\documentclass[journal]{IEEEtran}
\usepackage{graphicx}
\usepackage{amsmath}
\usepackage{algorithm}
\usepackage{hyperref}
\usepackage{booktabs} 
\usepackage{tabularx} 

\usepackage{booktabs} 
\usepackage{tabularx} 
\usepackage{caption}  
\usepackage{mathtools}
\usepackage{amssymb}
\usepackage{amsmath}
\usepackage{multirow}
\usepackage{tabularx} 
\usepackage{array}
\usepackage{makecell}
\usepackage{textgreek}
\usepackage{comment}
\usepackage{bm}
\usepackage{booktabs}
\usepackage{color}
\usepackage{tabularx}
\usepackage{algorithmicx}
\usepackage{algpseudocode}
\usepackage{subcaption}
\usepackage[normalem]{ulem}
\usepackage{xcolor}
\usepackage{colortbl} 
\usepackage{multirow} 
\usepackage{makecell} 

\algnewcommand\Input{\item[\textbf{Input:}]}  
\algnewcommand\Output{\item[\textbf{Output:}]}  
\usepackage{graphicx} 
\begin{document}

\title{A Location-Aware Hybrid Deep Learning Framework for Dynamic Near-Far Field Channel Estimation in Low-Altitude UAV Communications}

\author{ Wenli Yuan, Kan Yu,~\IEEEmembership{Member,~IEEE}, Xiaowu Liu, Kaixuan Li, Qixun Zhang,~\IEEEmembership{Member,~IEEE}, and Zhiyong Feng,~\IEEEmembership{Senior Member,~IEEE}
\thanks{This work is supported by the National Natural Science Foundation of China with Grant 62301076, Fundamental Research Funds for the Central Universities with Grant  24820232023YQTD01, National Natural Science Foundation of China with Grants 62341101 and 62321001, Beijing Municipal Natural Science Foundation with Grant L232003, and National Key Research and Development Program of China with Grant 2022YFB4300403.
}

\thanks{W. Yuan is with the School of Computer Science, Qufu Normal University, Rizhao, P.R. China. E-mail: Ywl1104@163.com;}
\thanks{K. Yu (\emph{the corresponding author}) is with the Key Laboratory of Universal Wireless Communications, Ministry of Education, Beijing University of Posts and Telecommunications, Beijing, 100876, P.R. China. E-mail: kanyu1108@126.com;}
\thanks{X. Liu is with the School of Computer Science, Qufu Normal University, Rizhao, P.R. China. E-mail: liuxw@qfnu.edu.cn;}
\thanks{Q. Zhang is with the Key Laboratory of Universal
Wireless Communications, Ministry of Education, Beijing University of Posts and Telecommunications, Beijing, 100876, P.R. China. E-mail: zhangqixun@bupt.edu.cn;}
\thanks{K. Li is with the School of Computer Science, Qufu Normal University, Rizhao, P.R. China. E-mail: lkx0311@126.com;}
\thanks{Z. Feng is with the Key Laboratory of Universal Wireless Communications, Ministry of Education, Beijing University of Posts and Telecommunications, Beijing, 100876, P.R. China. E-mail: fengzy@bupt.edu.cn.}
}

\markboth{IEEE Transactions on Mobile Computing,~Vol.~, No.~, 2025}%
{Shell \Baogui Huang{\textit{et al.}}: Shortest Link Scheduling Under SINR}
\maketitle
\begin{abstract}
In low-altitude UAV communications, accurate channel estimation remains challenging due to the dynamic nature of air-to-ground links, exacerbated by high node mobility and the use of large-scale antenna arrays, which introduce hybrid near- and far-field propagation conditions. While conventional estimation methods rely on far-field assumptions, they fail to capture the intricate channel variations in near-field scenarios and overlook valuable geometric priors such as real-time transceiver positions. To overcome these limitations, this paper introduces a unified channel estimation framework based on a location-aware hybrid deep learning architecture. The proposed model synergistically combines convolutional neural networks (CNNs) for spatial feature extraction, bidirectional long short-term memory (BiLSTM) networks for modeling temporal evolution, and a multi-head self-attention mechanism to enhance focus on discriminative channel components. Furthermore, real-time transmitter and receiver locations are embedded as geometric priors, improving sensitivity to distance under near-field spherical wavefronts and boosting model generalization. Extensive simulations validate the effectiveness of the proposed approach, showing that it outperforms existing benchmarks by a significant margin—achieving at least a 30.25\% reduction in normalized mean square error (NMSE) on average.

\end{abstract}

\IEEEpeerreviewmaketitle

\section{Introduction}\label{sec:introduction}
The unmanned aerial vehicle (UAV) communications have demonstrated significant potential in low-altitude economic applications. Their flexible deployment, rapid coverage establishment, and high-altitude line-of-sight (LoS) capabilities enable reliable communication links in scenarios where conventional infrastructure is unavailable or disrupted, such as in emergency response and remote area coverage \cite{zhao2019uav, yin2023air}. However, the radio propagation environment in low-altitude settings is highly complex, presenting substantial challenges for accurate channel estimation in UAV-assisted communication systems.

One major challenge stems from the high mobility of both UAVs and ground users, which causes rapid temporal variations in air-to-ground (A2G) channels \cite{byun2023channel}. Compounding this issue, the ongoing evolution from 5G massive multiple-input multiple-output (MIMO) to 6G extremely large-scale MIMO (XL-MIMO) systems has led to the deployment of large antenna arrays to enhance spectral efficiency and coverage. This expansion in array aperture results in the division of the electromagnetic field into distinct near-field and far-field regions, with the boundary defined by the Rayleigh distance \cite{cui2022channel, rodriguez2018frequency}. Crucially, the channel characteristics in these two regions differ fundamentally: far-field channels depend primarily on angle parameters, whereas near-field channels are influenced by both angle and distance \cite{yang2025near}. In low-altitude UAV communications, node mobility causes continuous variation in communication distances, leading to dynamic transitions between near-field and far-field propagations. Consequently, conventional channel estimation methods, which typically rely on far-field assumptions, often fail to provide accurate channel state information (CSI) in such heterogeneous and time-varying environments.

Traditional pilot-based approaches, including least squares (LS) and minimum mean square error (MMSE), perform adequately in static settings. While LS offers low computational complexity, it suffers from poor performance in high-noise and dynamic scenarios. MMSE achieves better accuracy \emph{but requires prior statistical knowledge of the channel and entails higher computational cost}. Both methods exhibit significant performance degradation under rapidly changing channel conditions \cite{liu2020overcoming}. Deep learning (DL) techniques have emerged as a promising alternative, capable of learning complex mappings from received signals to channel responses without explicit channel modeling or strong prior assumptions \cite{hu2020deep}. For instance, CNNs can effectively extract spatial features from received signals to capture local correlations among channel elements \cite{jiang2021dual}. Long short-term memory (LSTM) networks leverage temporal dependencies in sequential data to track channel variations over time \cite{nguyen2023channel}, while attention mechanisms help highlight semantically critical features for improved estimation performance \cite{kim2023towards}.

Nevertheless, most existing DL-based channel estimation methods focus predominantly on far-field scenarios, with limited consideration for the unique spatio-temporal evolution of channels in near-field environments. Moreover, they seldom incorporate geometric prior information—such as real-time transmitter and receiver positions—which could provide valuable inductive bias for distinguishing between near-field and far-field propagation modes. As a result, current approaches struggle to meet the accuracy and robustness requirements for channel estimation in practical UAV relay systems operating under dynamically near-far field propagation conditions.

this paper proposes a unified channel estimation framework suitable for both far-field and near-field regimes. The method integrates deep feature learning with location-assisted information to effectively capture the joint spatio-temporal evolution of the channel. The main contributions of this work are summarized as follows:

\begin{itemize}
     \item Unified channel estimation framework: We propose a unified channel estimation method for near- and far-field propagation which combines CNN and multi-head self-attention mechanism to capture local and global features, and employs BiLSTM to model bidirectional temporal dynamics, thereby enhancing modeling capability under multipath and dynamic channels.
     \item Position-aided prior fusion: We incorporate transceiver position information as geometric priors into the deep network, building a position-aided branch that is jointly modeled with learned features to enhance near-field distance sensitivity and bolster the robustness of channel estimation.
     \item Experimental validation in dynamic near-far field: Rigorous simulations compared to benchmarks demonstrate superior channel estimation accuracy and robustness. In particular, the framework achieves an average NMSE reduction of at least 30.25\% in near-far field scenarios under an SNR of 15dB.
 \end{itemize}

The remainder of this paper is organized as follows. Section \ref{sec:related work} reviews the latest developments in far-field and near-field channel estimation methods. In Section \ref{sec:network model}, the system model is presented, and the NMSE minimization problem is defined in channel estimation. Section \ref{sec:opt} provides a detailed introduction to the proposed unified deep learning-based channel estimation framework. Section \ref{sec:simulation} offers experimental evaluations of the proposed method. Finally, conclusions and directions for future work are provided in Section \ref{sec:conclusion}.

\section{Related Works}\label{sec:related work}
Channel estimation plays a pivotal role in wireless communication systems, directly impacting the signal transmission quality and overall performance. Accurate estimation enhances interference suppression and signal recovery, yet growing environmental complexity poses severe challenges to conventional methods. Recent advances in DL offer promising alternatives, transforming channel estimation into a data-driven learning task. In this section, we review state-of-the-art techniques, classified into methodologies based on far-field, near-field, and DL, with key contributions summarized in Table~\ref{tab:mobility_studies}.

\begin{table*}[t]
\caption{\small Representative Studies on Classical Channel Estimation Methods}
\label{tab:mobility_studies}
\centering
\arrayrulecolor{black}
\arrayrulewidth=0.5pt
\renewcommand{\arraystretch}{1.4}
\begin{tabular}{|
>{ \centering\arraybackslash}m{0.8cm}|
>{ \centering\arraybackslash}m{3.5cm}|
>{ \centering\arraybackslash}m{1.5cm}|
>{ \centering\arraybackslash}m{1.5cm}|
>{ \centering\arraybackslash}m{1cm}|
>{ \centering\arraybackslash}m{1.9cm}|
>{ \centering\arraybackslash}m{1.5cm}|
>{ \centering\arraybackslash}m{1.5cm}|
>{ \centering\arraybackslash}m{0.9cm}|
}
\hline
\textbf{Rref.} 
& \textbf{Techniques} 
& \textbf{Intelligence} 
& \textbf{
Channel type
}
& \textbf{Residual}
& \textbf{Attention}
& \textbf{Location information} 
&\textbf{NMSE/MSE} 
&\textbf{Mobility}\\ \hline

\cite{xu2022channel}
& MMSE
& $\times$
& Far-field
& $\times$ 
& $\times$ 
& $\times$ 
& $\checkmark$
& $\checkmark$
\\  \hline

\cite{zhu2025adaptive}
& Joint SBL and KF
& $\times$
& Near-field
& $\times$ 
& $\times$ 
& $\times$ 
& $\checkmark$
& $\checkmark$
\\  \hline

\cite{jang2024neural}
& Neural Network
& $\checkmark$
& Far-field and Near-field
& $\times$ 
& $\times$ 
& $\times$ 
& $\checkmark$
& $\times$
\\  \hline

\cite{yu2022deep}
& DNN and Stacked BiLSTM
& $\checkmark$
& Far-field
& $\times$ 
& $\times$ 
& $\times$ 
& $\checkmark$
& $\checkmark$
\\  \hline

\cite{nguyen2023channel}
& CNN and LSTM
& $\checkmark$
& Far-field
& $\times$ 
& $\times$ 
& $\times$ 
& $\times$
& $\checkmark$
\\  \hline

\cite{rahman2023hydnn}
& CNN and BiLSTM
& $\checkmark$
& Far-field
& $\times$ 
& $\times$ 
& $\times$ 
& $\times$
& $\times$
\\  \hline

\cite{liao2019deep}
& CNN and BiLSTM
& $\checkmark$
& Far-field
& $\times$ 
& $\times$ 
& $\times$ 
& $\checkmark$
& $\checkmark$
\\  \hline

\cite{zheng2025channel}
&  Deep Residual Attention Network
& $\checkmark$
& Far-field
& $\checkmark$ 
& SA and CA 
& $\times$ 
& $\checkmark$
& $\times$
\\  \hline

\cite{gao2022deep}
&  Attention Residual Network
& $\checkmark$
& Far-field
& $\checkmark$ 
& CA 
& $\times$ 
& $\checkmark$
& $\checkmark$
\\  \hline

\cite{chen2020channel}
& CNN and Transformer
& $\checkmark$
& Far-field
& $\checkmark$ 
& Muti-Head Self-Attention 
& $\times$ 
& $\checkmark$
& $\checkmark$
\\  \hline

\cite{guo2024parallel}
& Parallel Attention-Based Transformer
& $\checkmark$
& Far-field
& $\checkmark$
& Multi-Head Self-Attention
& $\times$ 
& $\checkmark$
& $\checkmark$
\\  \hline

\cite{lam2025racnn}
& RACNN
& $\checkmark$
& Far-field and Near-field
& $\checkmark$
& Self-Attention
& $\times$ 
& $\checkmark$
& $\times$
\\  \hline

\textbf{our work} 
& RACNN and BiLSTM 
& $\checkmark$ 
& Far-field and Near-field  
& $\checkmark$ 
& Multi-Head Self-Attention  
& $\checkmark$  
& $\checkmark$  
& $\checkmark$\\  \hline

\end{tabular}
\vspace{0.5cm}
\raggedright
\end{table*}

\subsection{Far-Field Channel Estimation Technologies}
Conventional far-field channel estimation techniques, such as LS and MMSE \cite{liu2020overcoming}, have been widely adopted. \cite{chen2023channel} and \cite{chung2024efficient} both exploited the sparsity of the shared base station (BS)-to-reconfigurable intelligent surface (RIS) channel and respectively proposed methods based on compressive sensing (CS) and low-rank matrix completion (LRMC) to jointly estimate the cascaded BS-RIS-user channel. Since RIS consists of passive reflecting elements and cannot actively process signals, channel estimation must be carried out by the BS, which consequently means that the BS-RIS and RIS-user channels cannot be directly estimated independently. Therefore, Zhou \emph{et al.} constructed a reciprocity link within a full-duplex BS and employed parallel factor (PARAFAC) analysis decomposition and atomic norm minimization (ANM) at the BS to effectively decouple and estimate the BS-RIS and RIS-user channels, respectively \cite{zhou2024individual}. Rekik \emph{et al.} exploited the orthogonality property of the orthogonal frequency division multiplexing (OFDM) system to independently estimate the covariance matrix and noise subspace for each subcarrier, and then minimized a global cost function in parallel to estimate channel coefficients efficiently~\cite{rekik2024fast}. In the MA systems, Xiao \emph{et al.} proposed a CS framework to recover the multi-path components (MPC) by exploiting the sparsity in the angular domain  \cite{xiao2024channel}.

A key limitation of the above-mentioned far-field studies is the neglect of time-varying channel characteristics under mobility. Xu \emph{et al.} addressed Doppler effects and doubly selective fading in high-mobility RIS systems using MMSE-based interpolation and reflection optimization~\cite{xu2022channel,xu2022reconfigurable}. Byun \emph{et al.} decomposed UAV-RIS channels into slow-varying angles and fast-varying gains, estimated at different timescales~\cite{byun2023channel}.

\subsection{Near-Field Channel Estimation Technologies}
The emergence of XL-MIMO and high-frequency communications has brought near-field propagation to the forefront, where traditional plane-wave assumptions break down~\cite{cui2022channel}. Recent studies address both channel estimation and user localization under spherical wavefronts. For instance, Yang \emph{et al.} and Lei \emph{et al.} systematically analyzed near-field characteristics—spherical waves, finite-depth beam focusing, and spatial non-stationarity—and reviewed corresponding localization and estimation algorithms~\cite{yang2025near,lei2025near}.

In \cite{lu2023near}, Lu \emph{et al.} proposed a hybrid LoS and non-line of sight (NLoS) near-field channel estimation model, separating the geometric free-space LoS path from NLoS components and introducing the MIMO advanced Rayleigh distance to define model boundaries. Cui \emph{et al.} introduced polar-domain sparse representations and developed both on-grid and off-grid recovery algorithms applicable to near- and far-field scenarios~\cite{cui2022channel}. Although the polar-domain channel representation method can characterize the spherical wave properties, it requires sampling in both the angle and distance domains, leading to high storage overhead and high dictionary coherence. To address these issues, Zhang \emph{et al.} proposed a distance-parameterized angular-domain model with a joint dictionary learning and orthogonal matching pursuit (OMP) algorithm. In
this approach, the user distance is treated as an unknown parameter in the dictionary, so that the number of dictionary columns depends only on the angular resolution, thereby avoiding the storage and computational burden~\cite{zhang2023near}.

For the channel estimation in wideband millimeter-wave XL-RIS systems, Yang \emph{et al.} analyzed the near-field beam squint effect and designed a wideband spherical-domain dictionary with angle-distance dimension sampling to minimize the atomic coherence \cite{yang2024near}. 
To address the issue that both CS and traditional bayesian learning (BL) methods struggle to simultaneously capture sub-array-specific sparsity and sparsity shared across sub-arrays, In \cite{pisharody2024near}, Pisharody \emph{et al.} modeled sub-array sparsity using Gaussian maxture priors and developed both centralized and decentralized variational BL algorithms. 
Jiang \emph{et al.} propose a neural network-aided joint optimization approach that simultaneously optimizes the beamformer and the localization function, effectively integrating data capturing environmental characteristics into the system design and further exploiting range information embedded in frequency selectivity to improve estimation accuracy \cite{jang2024neural}.

Few near-field studies adequately address mobility-induced time variation. Zhu \emph{et al.} proposed an adaptive joint sparse Bayesian Kalman filter that leverages temporal correlation in time-varying channels~\cite{zhu2025adaptive}. While in \cite{haghshenas2024parametric},  Haghshenas \emph{et al.} developed a parametric maximum likelihood framework with adaptive RIS configuration for channel estimation and user tracking under mobility and Rician fading.

\subsection{DL-Based Channel Estimation Technologies}
DL excels at learning nonlinear mappings and implicitly capturing structural dependencies, making it well-suited for channel estimation. Subsequently, Liu \emph{et al.} deeply integrated physical channel correlations with DL architectures, paving the way for numerous subsequent data-driven estimators~\cite{liu2020overcoming}. In \cite{mei2021performance}, Mei \emph{et al.} provided theoretical performance analysis for DL-based estimators, addressing a gap in earlier simulation-only validations. In \cite{gizzini2022survey} and \cite{gizzini2021cnn}, Gizzini \emph{et al.} surveyed DL methods for high-mobility and frequency-selective channels and proposed a CNN-based weighted interpolation estimator.

Hybrid architectures are widely adopted to capture both spatial and temporal features. CNN-LSTM models combine spatial feature extraction with temporal dependency modeling~\cite{nguyen2023channel,luo2018channel,cai2025deep}. BiLSTM variants further improve temporal modeling by incorporating both forward and backward contexts~\cite{rahman2023hydnn,liao2019deep,liao2019chanestnet}. In \cite{zhou2022deep}, Zhou \emph{et al.} combined CNN and convolutional LSTM for spatiotemporal channel prediction. In \cite{yuan2020machine}, Yuan \emph{et al.} integrated CNN with autoregressive and recurrent models for multi-step forecasting.

Deep neural network (DNN)--based approaches have also evolved. For example, in \cite{zheng2021simultaneous}, Zheng \emph{et al.} designed an adaptive online DNN trained solely on pilot signals, eliminating the need for true channel labels. In \cite{zhang2021designing}, Zhang \emph{et al.} introduced a tensor-train DNN to reduce parameter size and allow recursive estimation by incorporating historical information. In \cite{guo2023dnn}, Guo \emph{et al.} used DNNs to predict key channel parameters (such as gain, delay, and Doppler shift) rather than the full channel matrix. In \cite{yang2019deep}, Yang \emph{et al.} designed a DNN for doubly selective fading channels, incorporating pilots and historical CSI for tracking. In \cite{gao2023deep}, Gao \emph{et al.} combined sparse Bayesian learning with DNNs to mitigate power leakage and beam squint in mmWave MIMO systems. In \cite{yu2022deep}, Yu \emph{et al.} introduced a 3D geometric channel model for IRS-assisted UAV systems, combining DNN-based pre-estimation with BiLSTM-based tracking.

Attention mechanisms and residual connections further enhance the performance of DL-based channel estimators \cite{kim2023towards}, \cite{gao2022deep}. Residual connections alleviate gradient vanishing and enable deeper networks, while attention mechanisms explicitly model inter-feature dependencies. Specifically, channel attention (CA) and spatial attention (SA) reweight informative channels and locations, respectively. Transformer-based self-attention captures global and long-range dependencies through query-key-value interactions~\cite{kim2023towards}. As a result, several studies have successfully integrated these concepts into channel estimation frameworks. In \cite{guo2024parallel}, Guo \emph{et al.} reformulated the problem as a parallel super-resolution task using multi-head attention. In \cite{chen2020channel}, Chen \emph{et al.} combined CNN and Transformer to capture long-range and nonlinear temporal dynamics. In \cite{sun2023transformer}, Sun \emph{et al.} designed a lightweight Transformer for time-series refinement of channel estimates. In \cite{ju2024transformer}, Ju \emph{et al.} used multi-head self-attention to learn spatio-temporal correlations and combat channel aging. In \cite{lam2025racnn}, Lam \emph{et al.} proposed a residual attention CNN framework for hybrid-field channels. While Gao and Zheng \emph{et al.} integrated channel and spatial attention with residual learning for noise suppression and high-resolution reconstruction~\cite{zheng2025channel,gao2022deep}.

Unlike previous studies focused on static or single-scenario estimation, this work tackles time-varying channels in dynamic dual scenarios by incorporating real-time transceiver positions as geometric priors and jointly modeling spatio-temporal features within a unified learning framework.

Notations: In this paper, $(\cdot)^H$ and $(\cdot)^T$ denote the Hermitian (conjugate transpose) and transpose operations, respectively.

\begin{figure*}
\centering
\includegraphics[width=5.2in]{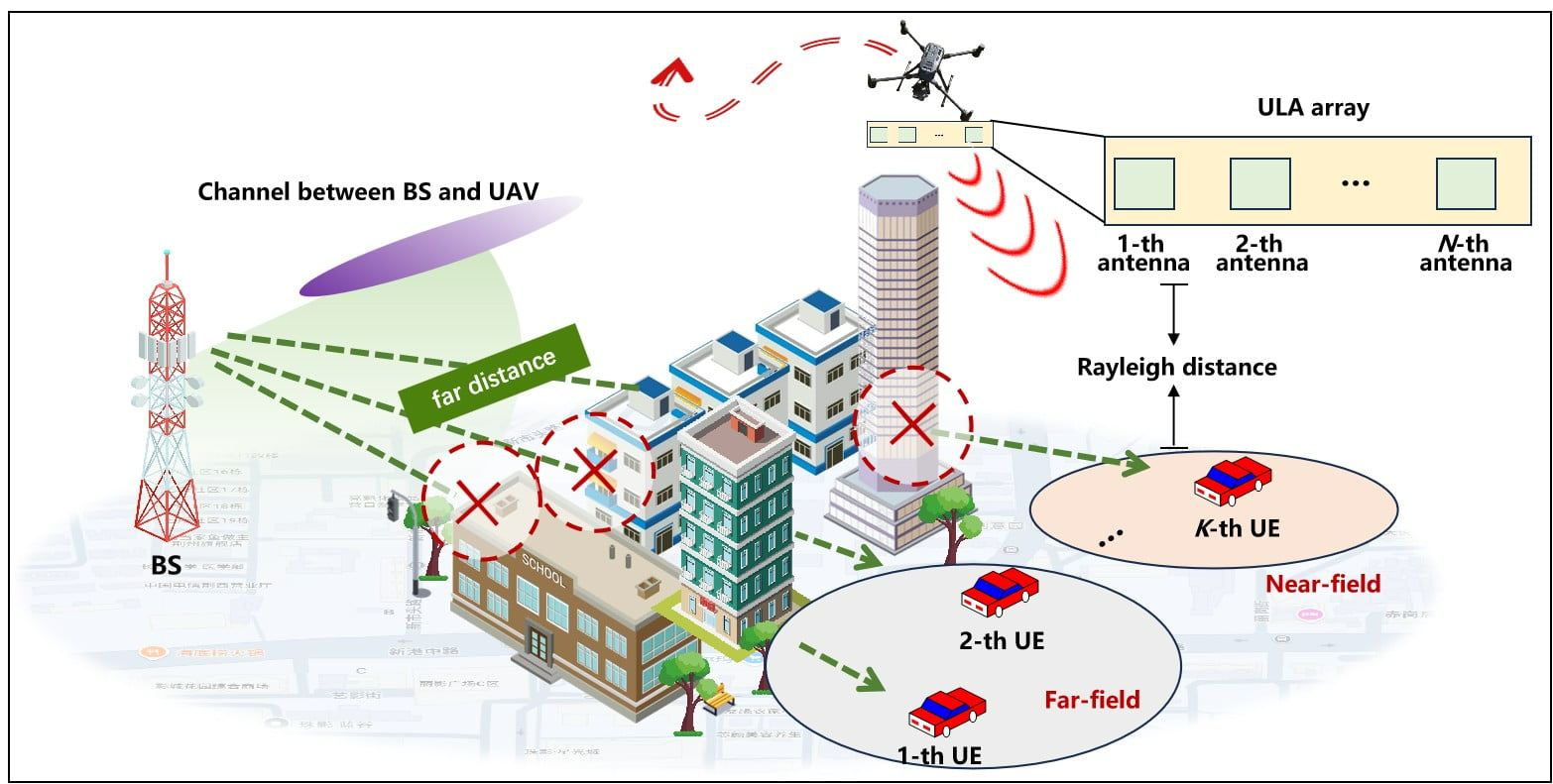}
\caption{\small Network model}
\label{fig:scenario}
\end{figure*}

\section{Network Model and Preliminaries}\label{sec:network model}
Consider an OFDM time division duplex (TDD) communication scenario, where the BS is unable to provide effective communication due to distance and obstructions. UAV serves as a relay system, equipped with $N$-element uniform linear array (ULA) and $N_{\mathrm{RF}}$ RF chains, providing reliable communication to $K$ ground user equipments (UEs). Due to the mobility of the UAV and the users, as well as environmental factors, both LOS and NLOS links exist in the communication channel. The UAV and UEs are assumed to move with velocities $v_U$ and $v_k$, respectively. As shown in Fig. \ref{fig:scenario}, in practical applications, UAV's flight altitude together with its relative position to the ground UE determines the propagation region—near field or far field \cite{cui2022channel}. Therefore, the relative distance between the UAV and the UE plays a significant role in the communication channel.


\subsection{Signal Model}
For uplink channel estimation, all users employ mutually orthogonal pilot sequences. This orthogonality allows the channel estimation for each user to be performed independently. We focus on an arbitrary user, who is assigned $M$ subcarriers with a pilot length of $Q$. In the $t$-th time slot, the UE transmits a known orthogonal pilot sequence, denoted by $s_m(t)$, on the $m$-th subcarrier. The corresponding received pilot signal at the UAV, denoted by $\mathbf{y}_m(t) \in \mathbb{C}^{N{\mathrm{RF}} \times 1}$, is then given by
\begin{equation}
  \mathbf{y}_m(t) = \mathbf{W}_{\mathrm{RF}}^{H}(t) \mathbf{h}_m s_m(t) + \mathbf{W}_{\mathrm{RF}}^{H}(t) \mathbf{n}_m(t),
\end{equation}
where $\mathbf{W}_{\mathrm{RF}}^{H}(t)\!\in\!\mathbb{C}^{N_{\mathrm{RF}}\times N}$ denotes the UAV RF combining matrix, 
$\mathbf{h}_m\!\in\!\mathbb{C}^{N\times 1}$ is the channel at the $m$-th subcarrier, and 
$\mathbf{n}_m(t)\!\sim\!\mathcal{CN}\!\left(\mathbf{0},\sigma^2 \mathbf{I}_N\right)$ is additive white Gaussian noise (AWGN). By concatenating all the received pilots at the $m$-th subcarrier (i.e., $\mathbf{y}_m =[\mathbf{y}_m^{T}(1),\,\mathbf{y}_m^{T}(2),\,\ldots,\,\mathbf{y}_m^{T}(Q)\big]^{T}
  \in \mathbb{C}^{Q N_{\mathrm{RF}}\times 1}$), we get
\begin{equation}
  \mathbf{y}_m = \mathbf{W}_{\mathrm{RF}}^{H} \mathbf{h}_m + \mathbf{n}_m,
\end{equation}
where $\mathbf{W}_{\mathrm{RF}} = \left[ \mathbf{W}_{\mathrm{RF}}(1), \ldots, \mathbf{W}_{\mathrm{RF}}(Q) \right] \in \mathbb{C}^{N \times Q N_{\mathrm{RF}}}$, and $\mathbf{n}_m = \left[ \mathbf{n}_m(1), \ldots, \mathbf{n}_m(Q) \right] \in \mathbb{C}^{Q N_{\mathrm{RF}} \times 1}$. Furthermore, the pilots received across all subcarriers (i.e. $\mathbf{Y} = \left[ \mathbf{y}_1, \ldots, \mathbf{y}_M \right] \in \mathbb{C}^{Q N_{\mathrm{RF}} \times M}
$) can be expressed as
\begin{equation}
\mathbf{Y} = \mathbf{W}_{\mathrm{RF}}^{H} \mathbf{H} \mathbf{S} + \mathbf{N},
\end{equation}
where $\mathbf{H} = \left[ \mathbf{h}_1, \ldots, \mathbf{h}_M \right] \in \mathbb{C}^{N \times M}
$, $\mathbf{S} = \left[ \mathbf{s}_1, \ldots, \mathbf{s}_M \right] \in \mathbb{C}^{N_{\mathrm{RF}} \times M}
$, and $\mathbf{N} = \left[ \mathbf{n}_1, \ldots, \mathbf{n}_M \right] \in \mathbb{C}^{Q N_{\mathrm{RF}} \times M}
$.
\subsection{Channel Model}

Conventional communication models predominantly rely on the far-field channel assumption, which approximates radiated electromagnetic waves as planar wavefronts \cite{cui2022channel}. However, this assumption becomes increasingly invalid with the growing number of antenna elements and reduced communication distances, where near-field effects are non-negligible \cite{zhang2023near}. The boundary between these regimes is defined by the Rayleigh distance, $d_F = 2D^2/\lambda$, where $D$ is the antenna array aperture and $\lambda$ is the wavelength \cite{cui2022channel}. In UAV communications, the link distance may fall on either side of $d_F$, leading to dynamic shifts in channel propagation characteristics. These far-field and near-field channels are fundamentally distinct, primarily in their dependencies on distance and angle, thus necessitating separate modeling approaches.

\subsection*{\textnormal{(1)} Far-field Channel Model}
When the UAV-UE separation exceeds the Rayleigh distance 
$d_F$, the channel operates in the far-field regime, characterized by a dependence on the angle of arrival (AoA) and a linear phase variation proportional to the propagation distance. The corresponding channel at the $m$-th subcarrier can be modeled as \cite{zhang2023near}
\begin{equation}
\mathbf{h}_m = \sqrt{\frac{N}{L}} \sum_{\ell=1}^{L} \alpha_{\ell} e^{-j \frac{2 \pi f_m r_{\ell}}{c}} \textnormal{a}(\theta_{\ell}),
\end{equation}
where $L$ is the number of effective propagation paths, $f_m$ is the carrier frequency, $c$ is the speed of light, $\alpha_{\ell}$ is the complex gain of the $l$-th propagation path, $r_{\ell}$ denotes the propagation distance from the center of the UAV array to the $\ell$-th propagation path, $\theta_{\ell}$ is the physical angle of the $\ell$-th path, and \( \textnormal{a}(\theta_{\ell}) \in \mathbb{C}^{N \times 1} \) represents the array response to the angle \( \theta_{\ell} \). For ULA, the far-field array response is represented as
\begin{equation}
\textnormal{a}(\theta_{\ell}) = \frac{1}{\sqrt{N}} \left[ 1, e^{j \frac{2 \pi d}{\lambda} \sin \theta_{\ell}}, \dots, e^{j \frac{2 \pi d}{\lambda} (N-1) \sin \theta_{\ell}} \right]^H,
\end{equation}
where $d=\lambda/2$ is the elements spacing of the antenna array.

\subsection*{\textnormal{(2)} Near-field Channel Model}

When the UAV-UE distance falls below the Rayleigh distance $d_F$, the incident wave demonstrates a spherical \cite{jang2024neural}. Consequently, both the amplitude and phase of the received signal exhibit significant variations across the antenna array due to the differing propagation distances. This necessitates an array response model that is jointly dependent on both the AoA and the distance. Thus, the near-field channel for the $m$-th subcarrier is modeled as \cite{zhu2025adaptive}
\begin{equation}
    \mathbf{h}_m = \sqrt{\frac{N}{L}} \sum_{\ell=1}^{L} \alpha_{\ell} e^{-j \frac{2 \pi f_m r_{\ell}}{c}} \textnormal{a}(\theta_{\ell}, r_{\ell}),
\end{equation}
where $\textnormal{a}(\theta_{\ell}, r_{\ell}) \in \mathbb{C}^{N \times 1}$ is the near-field array response vector for the ${\ell}$-th path, considering the combined effects of distance and angle. For ULA, the corresponding near-field array response vector is expressed as \cite{zhu2025adaptive}
\begin{equation}
    \textnormal{a}(\theta_{\ell}, r_{\ell}) = \frac{1}{\sqrt{N}} \left[ e^{j \frac{2 \pi}{\lambda} (r_{\ell}^{(0)} - r_{\ell})}, \ldots, e^{j \frac{2 \pi}{\lambda} (r_{\ell}^{(N-1)} - r_{\ell})} \right]^H,
\end{equation}
where $r_{\ell}^{(n)}$ is the distance between the UE and the $n$-th antenna, and $r_{\ell}$ denoted the distance between the center of the UAV array and UEs or scatters, which is expressed as
\begin{displaymath}
    r_{\ell}^{(n)} = \sqrt{r_{\ell}^2 + \delta_n^2 d^2 - 2 r_{\ell} \theta \delta_n d},
\end{displaymath}
where $\delta_n = \frac{2n - N - 1}{2}
$.
\subsection{Performance Metrics of Channel Estimation}
To quantitatively assess the quality of the proposed channel estimation framework, it is essential to define appropriate performance metrics. The mean squared error (MSE) directly measures the average squared discrepancy between the estimated and the true channel coefficients, providing a fundamental measure of estimation accuracy. To enable a fair comparison across different scenarios and channel conditions, this error is commonly normalized, leading to the normalized mean squared error (NMSE).

1) 
During training, the model parameters are optimized by minimizing the MSE between the channel estimates and the ground truth. To this end, the MSE loss function is expressed as follows \cite{cai2025deep}
\begin{equation}
\label{eq:MSE}
\mathcal{L}_{\mathrm{MSE}} = \|\mathbf{\hat{H}}_k - \mathbf{H}_k\|^2, 
\end{equation}
where $\mathbf{\hat{H}}_k$ represents the estimated channel, $\mathbf{H}_k$ is the true channel, and \( \| \cdot \|^2 \) denotes the the L2 norm.

2) 
The NMSE serves as a key performance metric for the channel estimation algorithm. A lower NMSE corresponds to higher estimation accuracy. This metric is adopted and defined as follows \cite{yu2022deep}
\begin{equation}
    \mathrm{NMSE} = \frac{\mathbb{E}\left( \| \mathbf{\hat{H}}_k - \mathbf{H}_k \|^2 \right)}{\mathbb{E}\left( \| \mathbf{H}_k \|^2 \right)}.
\end{equation}

\subsection{Problem Formulation}
Considering the scenario where UAV-assisted communication systems operate in low-altitude environments, with the distance between the UAV and users continuously varying, leading to dynamic switching between near-field and far-field channels, the goal is to use the observable received signal $\mathbf{Y}$ to estimate the near-field and far-field channels $\mathbf{H}$. For the $t$-th time slot, the propagation channel between the $k$-th UE and UAV is denoted as $\mathbf{H}_k(t)$, and the corresponding estimate is $\hat{\mathbf{H}}_k(t)$. The objective of network training is to minimize the MSE between the estimated channel and the true channel. Therefore, the pilot-aided channel estimation problem is formulated as
\begin{equation}
\label{eq:formulated problem}
\begin{aligned}
\min\;& \mathbb{E}\!\left[\big\| \mathbf{H}_k(t)-\hat{\mathbf{H}}_k(t)\big\|_{F}^{2}\right], \\
\text{s.t.}\;& \operatorname{tr}\!\left(\mathbf{s}_k \mathbf{s}_k^{H}\right)=E_s,
\end{aligned}
\end{equation}
where $\|\cdot\|_F^2$ denotes the squared Frobenius norm, $E_s = Q T_s P$ is the energy constraint, $T_s$ is the duration of a single time slot, and $P$ is the transmission power of the UAV. Conventional model-based estimators often struggle to capture the complex interplay between spatial variations, which are induced by large antenna arrays and varying propagation distances, and temporal dynamics resulting from the mobility of both UAVs and user equipment. To effectively address Problem \eqref{eq:formulated problem}, in the following section, we propose a DL-based solution that jointly exploits spatial and temporal channel structures.

\section{Proposed DL-Based Channel Estimation }\label{sec:opt}

\begin{figure*}
\centering
\includegraphics[width=6.5in]{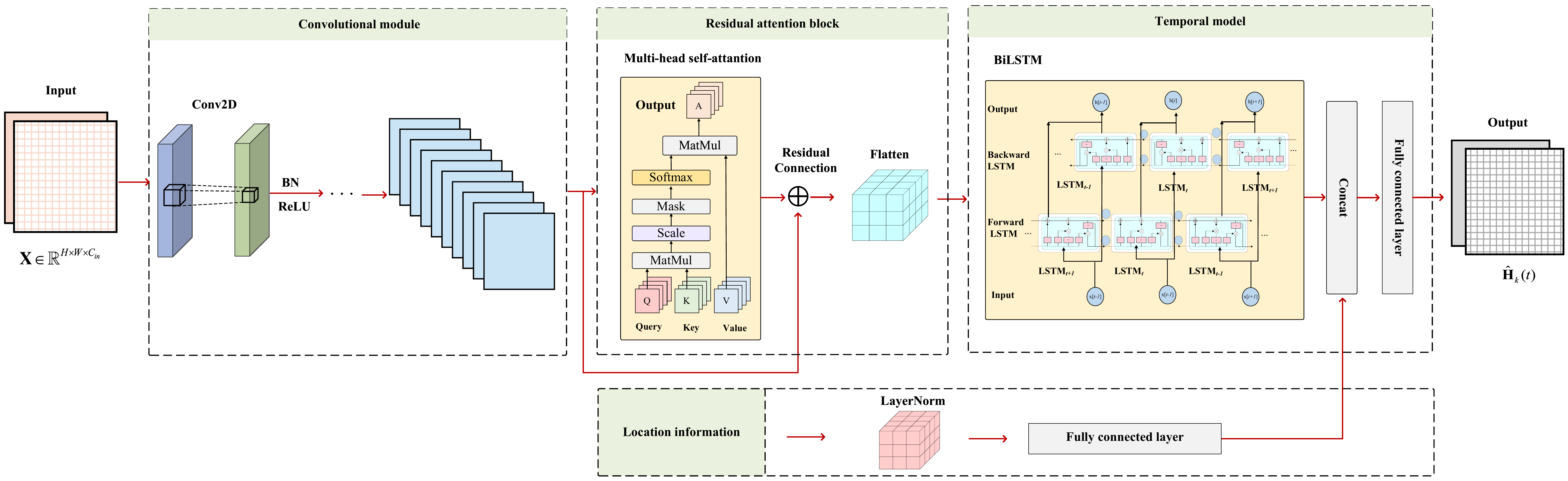}
\caption{\small The framework of the proposed channel estimation algorithm}
\label{fig:frame}
\end{figure*}

To address the challenge of channel estimation in dynamic propagation environments, in this section, we develop a novel DL method. The core architecture integrates three key components: CNNs for spatial feature extraction, a BiLSTM for capturing temporal evolution, and a residual attention mechanism for improved feature representation. 
This unified channel estimation framework, as illustrated in Fig. \ref{fig:frame}, facilitates the adaptive learning of complex channel characteristics across both far-field and near-field conditions.

\subsection{Data Preprocessing}

To formulate the channel estimation problem in vector form, we consider the signal matrix $\mathbf{Y}_k(t) \in \mathbb{C}^{N \times Q}$ received from the $k$-th UE in the $t$-th time slot. This matrix is vectorized into $\mathbf{y}_k(t) = \text{vec}(\mathbf{Y}_k(t)) \in \mathbb{C}^{N Q \times 1}$, and then split into real and imaginary components to form the real-valued input vector $\mathcal{Y}_k(t) = [\Re(\mathbf{y}_k(t))^T, \Im(\mathbf{y}_k(t))^T]^T \in \mathbb{R}^{2N Q}$. Similarly, the target channel matrix $\mathbf{H}_k(t)$ is vectorized as $\mathbf{h}_k(t)$, which serves as the regression target (or ``input image'' $\mathbf{X}$) for the DL model. A key aspect of our framework is that it does not require explicit prior knowledge of the near-field or far-field region. Instead, it learns to discern the underlying channel response patterns directly from the received signals, augmented by the real-time positional information of the UAV and the UE to enhance estimation accuracy. Specifically, given the known UAV position $\mathbf{r}_U(t) = (x_U(t), y_U(t), z_U(t))$ and UE position $\mathbf{r}_k(t) = (x_k(t), y_k(t), z_k(t))$ at the $t$-th time slot, the concatenated position vector $\mathbf{r} = [\mathbf{r}_U(t), \mathbf{r}_k(t)]$ is first normalized via Layer Normalization to stabilize training. It is then non-linearly projected by a fully-connected layer into a compact embedding, and can be given by
\begin{equation}
R=\sigma\!\left(\mathbf{W}\cdot\left(\gamma\,\frac{\mathrm{r}-\mu_{L}}{\sqrt{\sigma_{L}^{2}+\epsilon
}}+\beta\right)+b\right),
\end{equation}
where $\sigma(\cdot)$ is the ReLU activation function defined by $\sigma(x) = \max(0, x)$, $R$ denotes the processed position feature, $\mathbf{W}$ and $b$ are the weight matrix and bias term of the fully connected layer, $\gamma$, $\beta$, and $\epsilon$ are the learnable scale, shift parameters and a small constant for numerical stability. Assuming that the UAV and UE move with velocities $v_u$ and $v_k$ along fixed directions, their positions at each time step are updated as

\begin{displaymath}
\begin{aligned}
r_U(t) &= r_U(t-1) + v_U \cdot \Delta t \cdot d_U, \\
r_k(t) &= r_k(t-1) + v_k \cdot \Delta t \cdot d_k.
\end{aligned}
\end{displaymath}
where $d_U$ and $d_k$ denote the unit movement direction vectors of UAV and UE respectively,  $\Delta t$ is the time interval.

\subsection{Neural Network Architecture}

The proposed network model is architected around four interconnected components that function cohesively. (1) CNNs extract local spatial features from the channel observations, with its learning process stabilized by residual connections and its representational capacity enhanced by a Transformer attention mechanism, which directs the model's focus to critical regions. (2) The temporal dynamics inherent in the historical CSI, driven by the relative movement between the UAV and the UE, are captured by a BiLSTM network, thereby modeling the channel's evolution over time. (3) The real-time positional data of the UAV and the UE are embedded into the network as geometric priors, offering essential structural information about the propagation environment. (4) The high-level temporal features from the BiLSTM and the embedded geometric prior are concatenated and fused through a fully connected regression layer to produce the final estimate of the channel matrix for the current time step.

\subsubsection{\textbf{Convolutional-Attention Feature Extraction Module}}
This module consists of convolutional layers (conv2D), residual attention blocks (RA), ReLU activation function, and batch normalization (BN). We represent the image $\mathbf{X} \in \mathbb{R}^{H\times W\times C_{in}}$ as the input to the CNN module, where $H$, $W$, and $C_{\text{in}}$ denote the image height, image width, and the number of input channels, respectively. In the convolution operation, the data is convolved with the kernel $\mathrm{w}$, sliding over the input image. The stride used is set to $1$, and padding is applied to ensure that the convolution maintains the spatial dimension of the output image consistent with the input. We use several conv2D layers, and the corresponding mathematical expression is given by
\begin{equation}
\label{eq:Conv}
    \mathbf{X}'_{h,w,c} = \sum_{i=1}^{r} \sum_{j=1}^{r} \sum_{c'=1}^{C_{\text{in}}} \mathbf{X}_{h+i-1, w+j-1, c'} \cdot \mathrm{w}_{i,j,c',c} + b_c,
\end{equation}
where $\mathbf{X}'_{h,w,c}$ denotes the output of the convolution operation at position $(h,w)$ in the feature map $\mathbf{X}'$, the value at output channel $c$. $c'$ is the channel index of the input feature map, $i$ and $j$ are the sliding positions of the convolution kernel on the data. $\mathrm{w} \in \mathbb{R}^{r \times r \times C_{\text{in}} \times C_{\text{out}}}$ is the convolution kernel, with $r \times r$ representing the kernel size and $C_{\text{out}}$ representing the number of output channels. In our implementation, the first convolution uses $C_0 = 2$ as the input, and the number of output channels is set to $C_1 = 64$. In the subsequent layers, the number of input and output channels is fixed as $C_{\text{in}} = C_{\text{out}} = 64$. BN is applied after the convolutional layer for normalization. First, the feature map $\mathbf{X}'$ of the convolutional layer needs be normalized, denoted by $\hat{\mathbf{X}}'$, based on the conclusion of \cite{lam2025racnn}

\begin{equation}
\label{eq:norm}
    \hat{\mathbf{X}}' = \frac{\mathbf{X}' - \mu_B}{\sqrt{\sigma_B^2 + \epsilon'}},
\end{equation}
where $\mu_B$ and $\sigma_B^2$ are the mean and variance of the current batch data, and $\epsilon'$ is a small constant to prevent division by zero. Next, to restore the network's expressive capacity, scaling and shifting are typically applied to obtain the final BN output

\begin{equation}
\label{eq:BNout}
    \mathbf{X}_{\rm{BN}} = \gamma \hat{\mathbf{X}}' + \beta',
\end{equation}
where $\gamma'$ and $\beta'$ denote learnable scale and shift parameters, respectively. To reduce overfitting during the training process, the output of some Conv2D layers uses the ReLU activation function to introduce non-linearity. The output after the series of convolution, BN, and ReLU operations is represented as $\mathbf{X}_1$, which serves as the input to the subsequent residual attention block. The residual connection mitigates the gradient vanishing problem in deep networks while preserving the integrity of the original information, thus stabilizing the training process. The attention mechanism enables the network to adaptively focus on the most important regions of the input feature map, thereby highlighting the key features that are most valuable for channel estimation. Based on the Transformer multi-head attention architecture, we design a mechanism that allocates weights through the Query-Key-Value dot-product calculation (ranging from 0 to 1). These weights dynamically determine which relevant elements of the input sequence each output position should focus on, thereby establishing global dependencies and enhancing the model's ability to represent complex channel environments. Specifically, the input feature matrix $\mathbf{X}_1$ passes through three independent linear projection layers to generate the query matrix $\mathbf{Q}$, key matrix $\mathbf{K}$, and value matrix $\mathbf{V}$. When the number of attention heads is $N_h$, for each head $i$, the following calculations are performed \cite{guo2024parallel} 
\begin{equation}
\label{eq:matrix}
    \mathbf{Q}_i = \mathbf{X}_1 \mathbf{W}_i^q, \quad \mathbf{K}_i = \mathbf{X}_1 \mathbf{W}_i^k,~ \text{and}~\mathbf{V}_i = \mathbf{X}_1 \mathbf{W}_i^v,
\end{equation}
where $\mathbf{Q}_i$, $\mathbf{K}_i$, and $\mathbf{V}_i$ represent the corresponding $i$-th transformation matrices, $i \in \{1, 2, \dots, N_h\}$. The attention weights are computed by comparing the query and key vectors, where the difference between these two vectors determines the extent to which an input element should attend to other elements. Multi-head attention leverages the query $\mathbf{Q}_i$ and key $\mathbf{K}_i$ to capture the spatiotemporal dependencies between elements in the channel sequence. The scaled dot-product attention for each head $i$ is computed as
\begin{equation}
\label{eq:Attention}
    \mathbf{A}_i = \text{softmax}\left( \frac{\mathbf{Q}_i \mathbf{K}_i^T}{\sqrt{d_k}} \right) \mathbf{V}_i,
\end{equation}
where $d_k$ is the dimension of the key vector, $\sqrt{d_k}$ is the scaling factor to prevent the dot product result from becoming too large, which could lead to the softmax gradient vanishing. Then, each head performs a weighted sum of the value matrix $\mathbf{V}$ using the computed attention weight $\mathbf{A}_i$, and the output for each head can be calculated by $\text{head}_i= \mathbf{A}_i \times \mathbf{V}$. Multi-head attention enables the model to simultaneously focus on different subspaces of the input feature representations and process them in parallel. The outputs of all heads are concatenated and then mapped through a linear transformation to obtain the final output. That is,
\begin{equation}
\label{eq:Concat}
    \mathbf{X}_{\text{att}} = \text{Concat}(\text{head}_1, \text{head}_2, \dots, \text{head}_{N_h}) \mathbf{W}^O,
\end{equation}
where $\text{Concat}(\cdot)$ performs concatenation, and $\mathbf{W}^O$ is a weight matrix that projects the concatenated heads into a unified representation. This output is then additively combined with the input via a residual connection, followed by layer normalization. This process alleviates the problems of gradient vanishing/exploding and enhances training stability, producing the final normalized output, namely

\begin{equation}
\label{eq:res}
    \mathbf{X}_{\text{out}} = \text{LayerNorm}(\mathbf{X}_1 + \mathbf{X}_\text{{att}}),
\end{equation}
For temporal modeling, the spatial features are vectorized and input into the LSTM in time-major order to learn dependencies across time steps \cite{cai2025deep}.

\subsubsection{\textbf{LSTM Temporal Modeling Module}}
LSTM is an advanced variant of recurrent neural networks (RNN) that addresses the vanishing gradient and long-term dependency issues in traditional RNNs by incorporating gating mechanisms. These gates enable LSTM to selectively retain important temporal information and suppress irrelevant data, making it effective for time-series modeling. In our channel estimation framework, LSTM acts as the second module, receiving spatial features extracted by the CNN module for temporal modeling. The LSTM unit comprises three key gates: the forget gate, input gate, and output gate, as well as a cell state for long-term memory. At each time step, the gates determine which information to keep, update, or output based on the current input and the previous hidden state, ensuring efficient temporal dependency learning. The first layer is the forget gate, which is composed of information from the previous layer $x_{t-1}$ and the current $x_t$. It controls the retention of the previous time step's memory state $c_{t-1}$ at the current time step
\begin{equation}
\label{eq:f}
    f_{g,t} = \sigma(\mathbf{W}_f x_t + \mathbf{W}_f' x_{t-1} + b_f),
\end{equation}
The second layer is the input gate, which determines the degree of updating of the new information at the current time step
\begin{equation}
\label{eq:i}
    i_{g,t} = \sigma(\mathbf{W}_i x_t + \mathbf{W}_i' x_{t-1} + b_i),
\end{equation}
The third layer is the cell input state and further implemented by using the $\text{tanh}$ function
\begin{equation}
\label{eq:c}
    c_{g,t} = \text{tanh}(\mathbf{W}_c x_t + \mathbf{W}_c' x_{t-1} + b_c),
\end{equation}
The last layer is the output gate
\begin{equation}
\label{eq:o}
   o_{g,t} = \sigma(\mathbf{W}_o x_t + \mathbf{W}_o' x_{t-1} + b_o),   
\end{equation}
where $\mathbf{W}_f$, $\mathbf{W}_i$, $\mathbf{W}_c$, and $\mathbf{W}_o$ denote the input weights for each gate at the current time step.
$\mathbf{W}'_f$, $\mathbf{W}'_i$, $\mathbf{W}'_c$, and $\mathbf{W}'_o$ denote the historical state weights. $b_f$, $b_i$, ${b}_c$, and $b_o$ are the bias. $f_{g,t}$, $i_{g,t}$, and $o_{g,t}$ are control vectors. $\sigma(\cdot)$ is the gate activation function, which is typically chosen as the sigmoid. Apart from the candidate cell input state $c_{g,t}$ in the hidden layer, the architecture maintains two additional cell states: the previous cell state $c(t\!-\!1)$ received by the current LSTM unit and the current cell state $c(t)$ passes to the next LSTM unit. The output state at the current time step is updated as
\begin{equation}
\label{eq:C_t}
    c(t) = f_{g,t} \odot c(t-1) + i_{g,t} \odot c_{g,t},
\end{equation}
Finally, the output layer is computed as
\begin{equation}
\label{eq:output}
    h_t = o_{g,t} \odot \text{tanh}\!\big(c(t)\big),
\end{equation}
where the operation of $\odot$ represents the hadamard product. BiLSTM overcomes the limitation of LSTM that captures only past information by simultaneously integrating forward and backward information flows, thereby modeling both historical context and future tendencies. Specifically, a BiLSTM consists of two LSTM networks that process the input sequence in the forward and reverse directions, respectively, enabling a more comprehensive capture of temporal dependencies. Hence the forward output $\overrightarrow{h_t}$ and backward output $\overleftarrow{h_t}$ of each LSTM cell is calculated based on the relative input and the output layer function in Eq. \eqref{eq:output}. The output of the BiLSTM network at each time step can be represented as
\begin{equation}
\label{eq:Concat()}
    h_t = \text{Concat}(\overrightarrow{h_t}, \overleftarrow{h_t}),
\end{equation}
To fully leverage the temporal features and location information, the temporal features output by BiLSTM are concatenated with the location features obtained from the multilayer perceptron (MLP) mapping along the feature dimension to obtain 
\begin{equation}
\label{eq:R}
    h'_t = \text{Concat}(h_t, R).
\end{equation}
Subsequently, to align the network's output with the target channel dimensions, the concatenated feature is processed by a fully-connected layer that produces a real-valued vector of appropriate size. This vector is subsequently interpreted as containing the concatenated real and imaginary parts of the channel coefficients, thereby yielding the complex-valued estimate $\hat{\mathbf{h}}_k(t)$. The final channel matrix $\hat{\mathbf{H}}_k(t)$ is obtained by applying the requisite reshape operation to $\hat{\mathbf{h}}_k(t)$, dentoed by $\hat{\mathbf{H}}_k(t) = \text{reshape}\left( \hat{\mathbf{h}}_k(t) \right)$. The detailed steps of the proposed channel estimation algorithm are outlined in Algorithm \ref{Alg:algorithm1}.

\begin{algorithm}
\caption{Residual Attention-based CNN-BiLSTM Channel Estimation}
\label{Alg:algorithm1}
\begin{algorithmic}[1]
\small
\Statex \textbf{Module One CNN Feature Extraction:}
\Statex \textbf{Input:} Training received signal $\mathcal{Y}_k(1), \dots, \mathcal{Y}_k(T)$, training true channel information $\mathbf{H}_k(1), \dots, \mathbf{H}_k(T)$
\Statex \textbf{Initialization:} Randomizing initial weights $\mathbf{W}$ and bias $b$
\For{$t=1$ \textbf{to} $T$}
    \State Convolution processing with Eq. \eqref{eq:Conv}-Eq. \eqref{eq:BNout} and ReLU
    \For{each head $i \in \{1, 2, \dots, N_h\}$}
    \State Calculating $\mathbf{Q}_i$, $\mathbf{K}_i$, $\mathbf{V}_i$ with Eq. \eqref{eq:matrix}
    \State Calculating $\mathbf{A}_i$ with Eq. \eqref{eq:Attention} 
    \State Calculating $\mathbf{X}_\text{att}$ with Eq. \eqref{eq:Concat}
\EndFor
\State Residual connection with Eq. \eqref{eq:res}
\State Flatten
\EndFor
\Statex \textbf{Module Two LSTM Temporal Model:}
\Statex \textbf{Initialization:} Randomizing initial weights $\mathbf{W}$ and bias $b$
\State Temporal encoding over $t = 1,\dots,T$
\State Updating gates and states per Eq. \eqref{eq:f}-Eq. \eqref{eq:output} in forward and backward passes; concatenate per Eq. \eqref{eq:Concat()}
\State Concatenating position information with Eq. \eqref{eq:R}
\State \textbf{while} not convergence \textbf{do}
\State \quad Updating weights $\mathbf{W}$ and bias $b$ by minimizing the MSE loss function with Eq. \eqref{eq:MSE}
\State \textbf{end while}
\end{algorithmic}
\end{algorithm}

\textbf{Model training:}  
The proposed channel estimation model is trained under a supervised learning paradigm. The gradients of the loss function with respect to the network parameters are computed via backpropagation. For parameter optimization, we employ the Adam optimizer, which utilizes adaptive estimates of first- and second-order moments to facilitate stable and efficient convergence toward a high-quality solution.

  
\subsection{Complexity Analysis}
A computational complexity analysis of the proposed model reveals that the primary costs originate from the CNN, multi-head self-attention, BiLSTM, and fully-connected layers. The initial stage involves processing the input through convolutional layers to project it into a higher-dimensional feature space. The complexity of this convolutional stage is $\mathcal{O}(L_{\text{conv}} \cdot B \cdot H \cdot W \cdot C_{\text{in}} \cdot C_{\text{out}} \cdot k^2)$, where $B$ is the batch size, $L_{\text{conv}}$ is the number of convolutional layers, $H$ and $W$ denote the spatial dimensions of the feature maps, $C_{\text{in}}$ and $C_{\text{out}}$ are the input and output channels, and $k$ represents the kernel size. The self-attention mechanism in the Transformer module is computationally intensive, with a complexity of $\mathcal{O}(B \cdot T^2 \cdot d_{\text{model}})$, where $T$ is the sequence length and $d_{\text{model}}$ is the model dimension. The BiLSTM's complexity is linear with respect to the sequence length $T$ and quadratic with respect to the hidden state dimension $d_h$, yielding $\mathcal{O}(B \cdot T \cdot d_h^2)$. Finally, the total complexity of the fully-connected layers is given by $\sum_{i=1}^{L_{\text{dense}}} \mathcal{O}(B \cdot d_{\text{in}}^{(i)} \cdot d_{\text{out}}^{(i)})$, where $L_{\text{dense}}$ is the number of layers, and $d_{\text{in}}^{(i)}$ and $d_{\text{out}}^{(i)}$ denote the input and output dimensions of the $i$-th layer, respectively.

\section{Simulation Results}\label{sec:simulation}

In this section, we present numerical simulations to evaluate the performance of the proposed channel estimation scheme. The simulation encompasses both near-field and far-field scenarios, with the UAV and UE following predefined trajectories. The dataset is randomly split into an 80\% training set and a 20\% validation set. Key simulation parameters are listed in Table \ref{tab:sim para validation}.

\begin{table}[!htb]  \caption{\small Simulated parameters and values}
\centering
\label{tab:sim para validation}
\begin{tabular}{lll}
\toprule
  Symbol & Meanings &Values\\
\midrule
  $N$         & the number of antennas  & 16 \\
  $f$          &frequency band of communication & 3.5GHz\\
  $\lambda $   &  beamwave   & 0.0857m \\
  $d_F$        &  Rayleigh distance     & 9.6m    \\
  $Q$   & pilot length  & 8 \cite{jang2024neural}\\
  $v_U$     & the speed of UAV   & $5 \, \text{m/s}$ \\
  $v_k$     & the speed of UE   & $5 \, \text{m/s}$ \\
  $L_f$  &number of far-field paths   &3\\
  $L_n$   &number of near-field paths   &3\\
  $\eta$ & learning rate & 0.001\\
  \bottomrule
\end{tabular}
\end{table}

\subsection{Baseline frameworks}
To benchmark the performance of our proposed convolutional attention-based BiLSTM network, we conduct a comprehensive comparison with several established DL channel estimators. To this end, the following baseline methods are selected for performance evaluation:
\begin{itemize}
\item \textbf{CNN \cite{jiang2021dual}:} A classic CNN that extracts spatial features from channel data through convolution operations, enabling the model to capture spatial information from the input signal.
\item \textbf{RACNN \cite{lam2025racnn}:} CNN combines residual connections with the Transformer self-attention mechanism. It focuses on key regions within the feature map, enhancing the model’s ability to attend to important information.
\item \textbf{CNN-LSTM \cite{nguyen2023channel}:} This model combines the spatial feature extraction capability of CNN with the temporal modeling strength of LSTM. It first extracts spatial features from the input signal using CNN, then captures temporal dependencies in the signal using LSTM.
\item \textbf{DNN-BiLSTM \cite{yu2022deep}:} This model leverages the strengths of DNN and BiLSTM. DNN performs deep learning of nonlinear features through multiple fully connected layers, while BiLSTM captures both forward and backward temporal dependencies of the signal.
\item \textbf{CBAM-BiLSTM \cite{zheng2025channel}:} A model combining the CBAM attention mechanism with BiLSTM. CBAM is a lightweight attention mechanism that improves feature representation through SA and CA, which are then used in conjunction with BiLSTM to model temporal dependencies.
\item \textbf{RACNN-BiLSTM:} The proposed fusion architecture of convolutional attention and BiLSTM, which incorporates position information as auxiliary input. The residual CNN integrates the Transformer multi-head self-attention mechanism to enhance spatial feature representation, and BiLSTM is employed to model the temporal evolution of the channel.
\end{itemize}

\subsection{Performance evaluation}

\begin{figure}[!ht]
\centering
\includegraphics[width=1\linewidth]{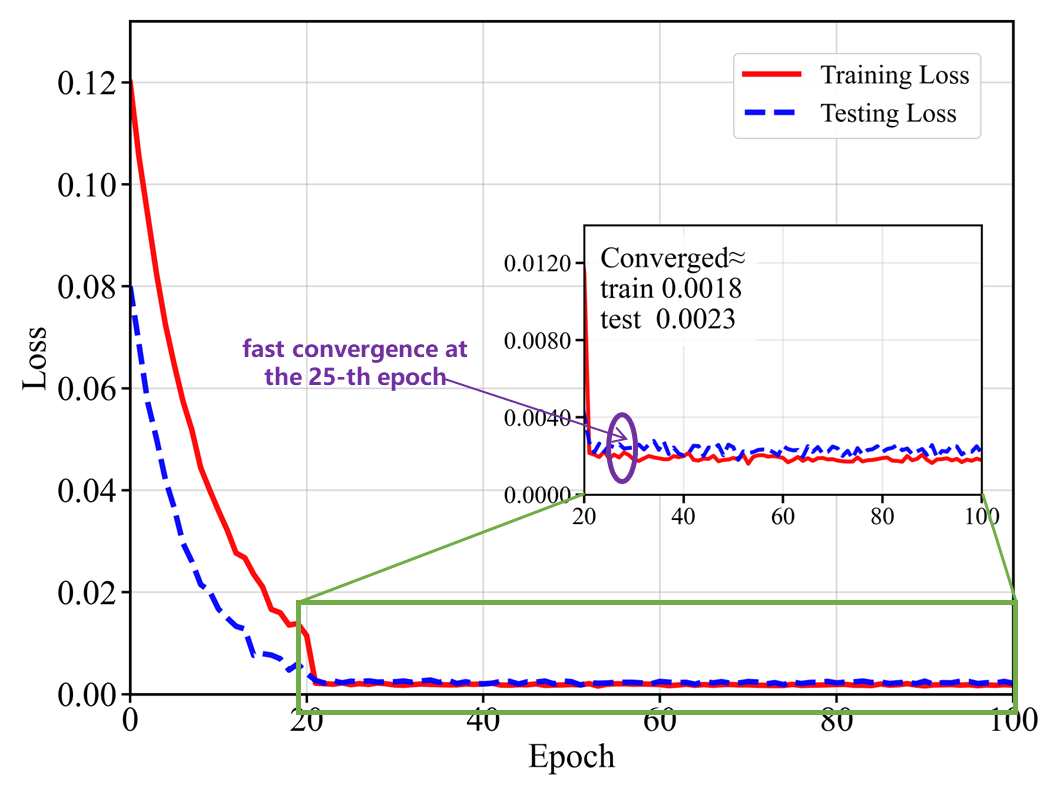}
\caption{\small RACNN-BiLSTM training vs. testing Loss}
\label{fig:training_loss}
\end{figure}

Fig. \ref{fig:training_loss} illustrates the variation of training loss and testing loss with respect to the number of training epochs during the training process of the proposed RACNN-BiLSTM method. As shown in the figure, both the training loss and test loss decrease rapidly in the early stages, indicating that the model successfully learns effective features in the initial phases. After approximately 25 epochs, the training loss stabilizes and remains at a relatively low level. Simultaneously, the test loss follows a similar trend, gradually converging and remaining consistent with the training loss, suggesting that the model does not experience overfitting and is able to maintain strong performance on the test set. Overall, the model demonstrates good convergence behavior, achieving low loss values in a relatively short training period, which highlights the effectiveness and stability of the proposed method.

\begin{figure}[!ht]
    \centering
    \begin{subfigure}[b]{3.5in}
        \centering
        \includegraphics[width=1\linewidth]{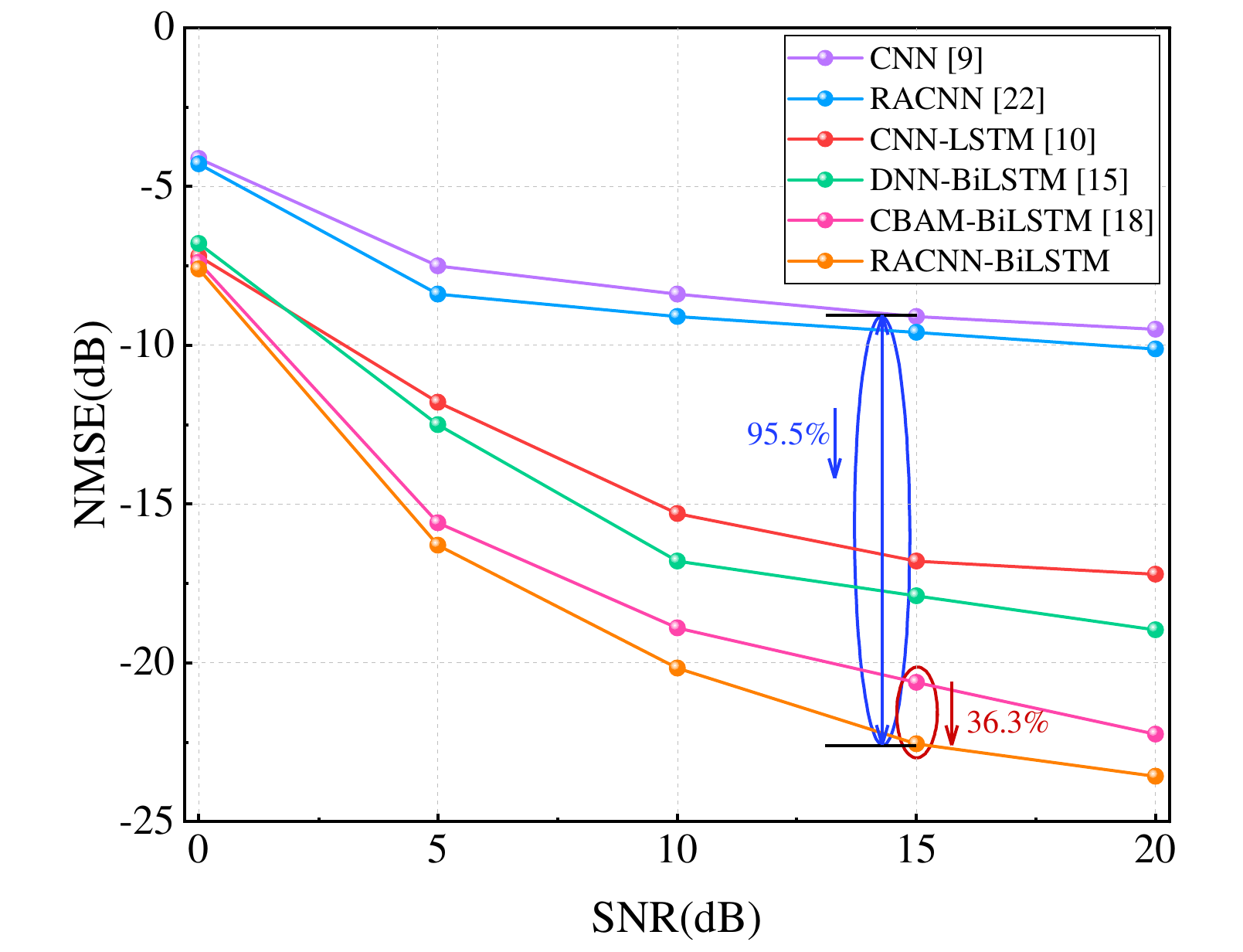}
        \caption{far-field: NMSE vs. SNR}
        \label{fig:far_field}
    \end{subfigure}
    
    \vspace{0.3cm}
    
    \begin{subfigure}[b]{3.5in}
        \centering
        \includegraphics[width=1\linewidth]{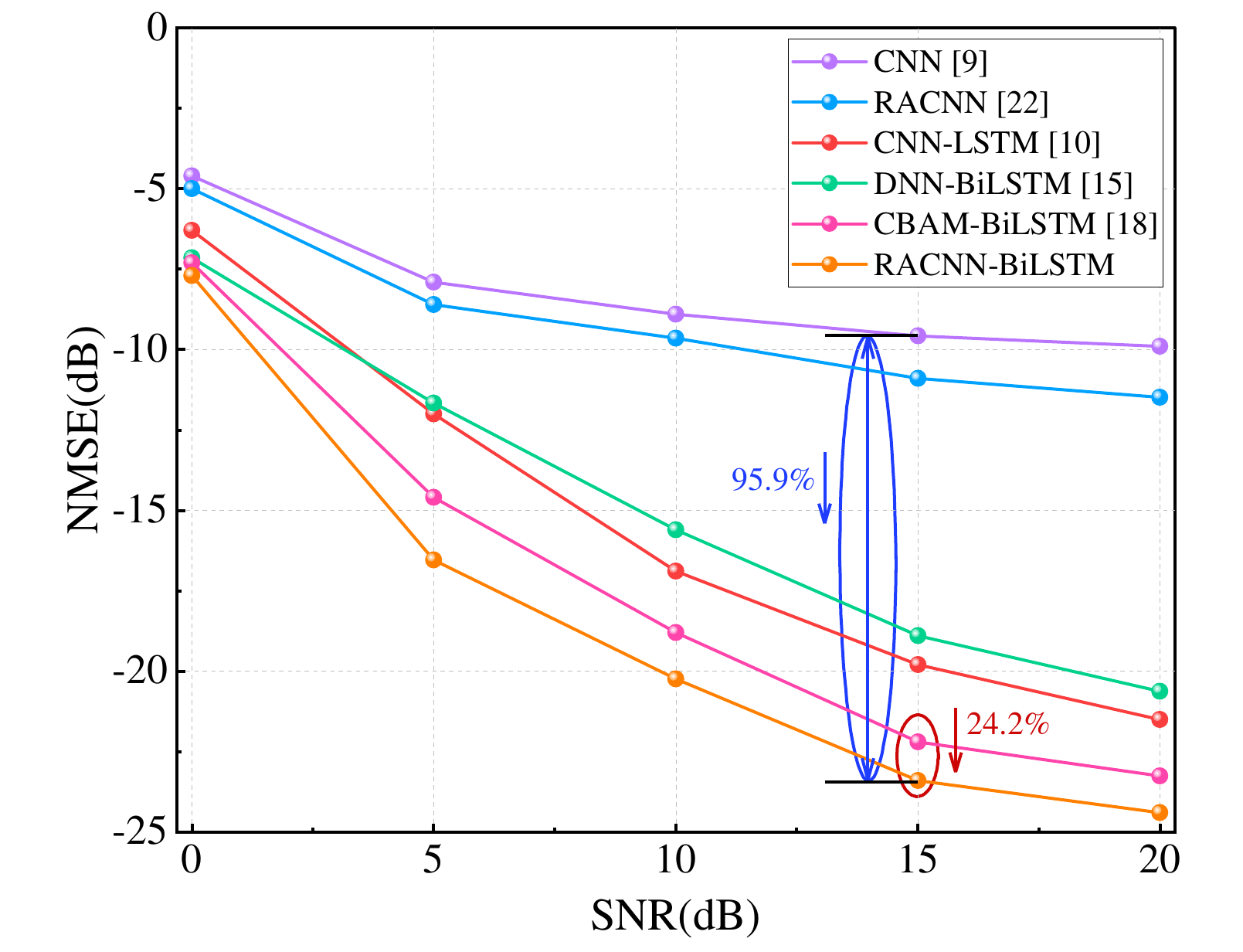}
        \caption{near-field: NMSE vs. SNR}
        \label{fig:near_field}
    \end{subfigure}
    
    \caption{\small NMSE vs. SNR for far-field and near-field.}
    \label{fig:snr_nmse}
\end{figure}

To evaluate the estimation performance of the proposed method, Fig. \ref{fig:snr_nmse} shows the NMSE performance curves of different algorithms under varying SNR conditions for both far-field and near-field scenarios. Under low SNR conditions, due to strong noise interference and low signal quality, the performance of all algorithms is poor. As the SNR increases, the NMSE performance of all algorithms decreases, indicating that improving the SNR effectively mitigates noise interference, thereby enhancing the accuracy of channel estimation. The proposed RACNN-BiLSTM algorithm consistently exhibits superior performance in both scenarios. The framework achieves an average NMSE reduction of at least 30.25\% in near-field and far-field scenarios under an SNR of 15 dB. In contrast, while RACNN also employs a residual attention mechanism, it lacks temporal modeling capability. The performance of CBAM-BiLSTM is inferior to that of RACNN-BiLSTM, primarily due to the differences in the attention mechanisms used and the lack of position-aided information. CNN, CNN-LSTM, and DNN-BiLSTM, which do not incorporate attention or residual structures, rely primarily on convolutional or fully connected networks for feature extraction. These models struggle to effectively capture key information in complex multipath and spatially correlated environments. Especially in near-field channels, where the spherical wave effects and spatial variations are more intricate, models lacking attention mechanisms fail to accurately capture spatial correlations, leading to inferior performance.

\begin{figure}[!ht]
\centering
\includegraphics[width=0.9\linewidth]{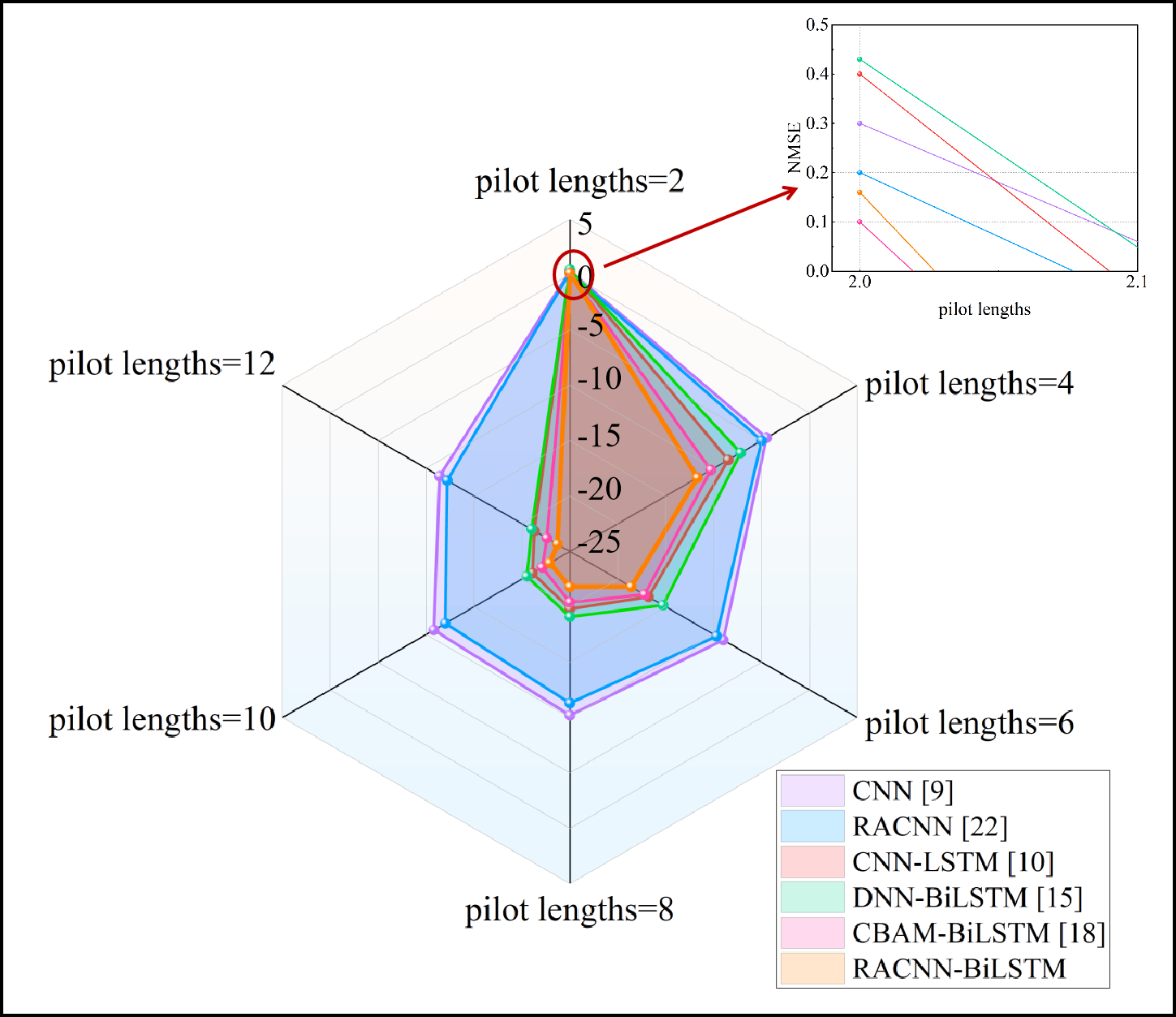}
\caption{\small near-field: NMSE vs. pilot lengths}
\label{fig:pilot}
\end{figure}

Fig. \ref{fig:pilot} illustrates the impact of different pilot lengths on NMSE performance in the near-field scenario with SNR of 10 dB. When the pilot length is 2, the channel estimation accuracy is limited by the scarcity of pilot information, resulting in relatively high NMSE values for all models. In this case, although the RACNN-BiLSTM model possesses strong spatiotemporal modeling capabilities, it cannot fully leverage its advantages due to the lack of sufficient pilot data. As the pilot length increases, the NMSE values of all models gradually decrease, indicating a significant improvement in channel estimation accuracy. Notably, the RACNN-BiLSTM model consistently maintains the lowest NMSE values across the higher pilot length range, demonstrating superior channel estimation performance compared to other methods.

\begin{figure*}[!ht]
    \centering
    \begin{subfigure}{0.32\textwidth}
        \centering
        \includegraphics[width=1\linewidth]{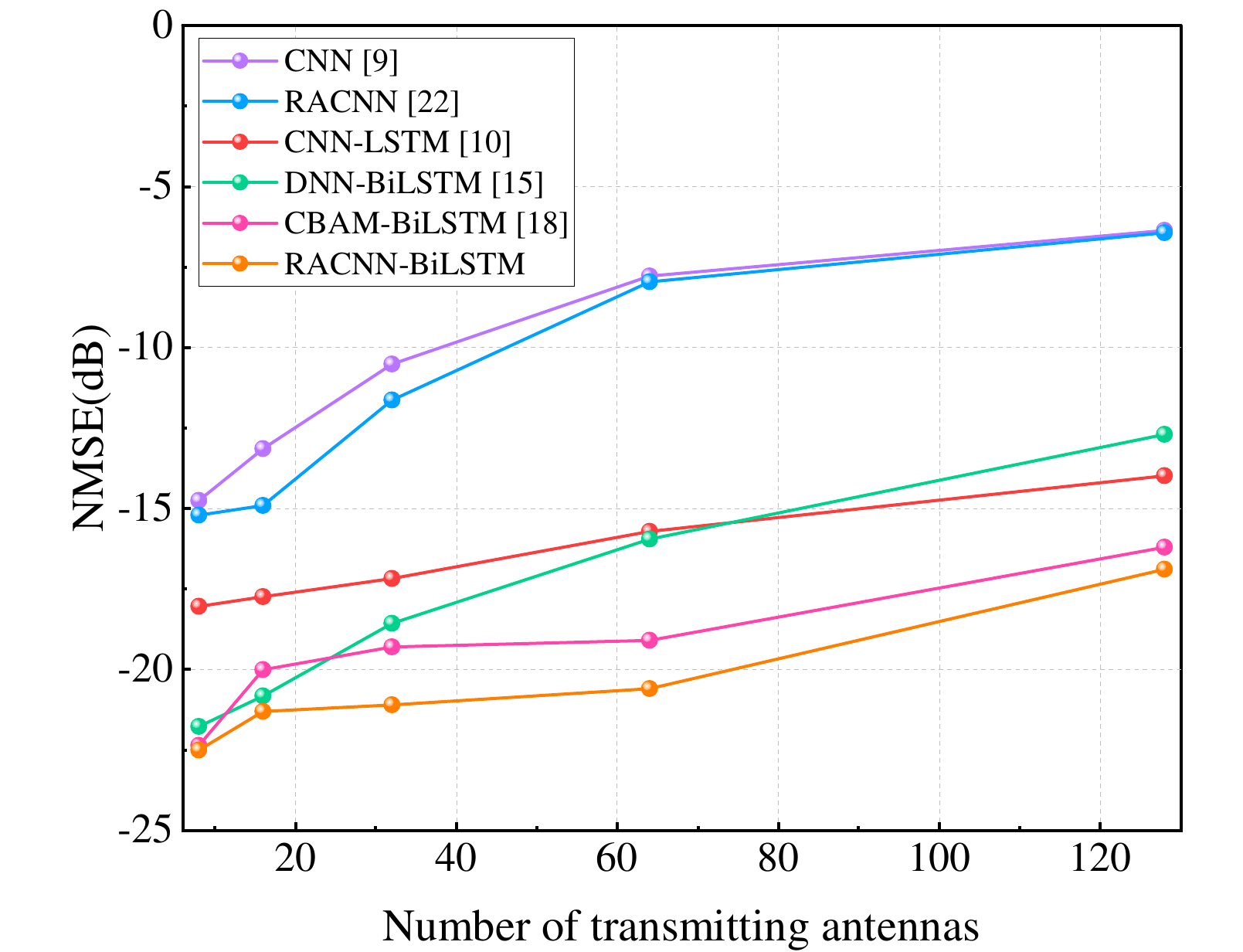}
        \caption{\small Number of transmitting antennas }
        \label{fig:Nt_far}
    \end{subfigure}
    \begin{subfigure}{0.32\textwidth}
        \centering
        \includegraphics[width=1\linewidth]{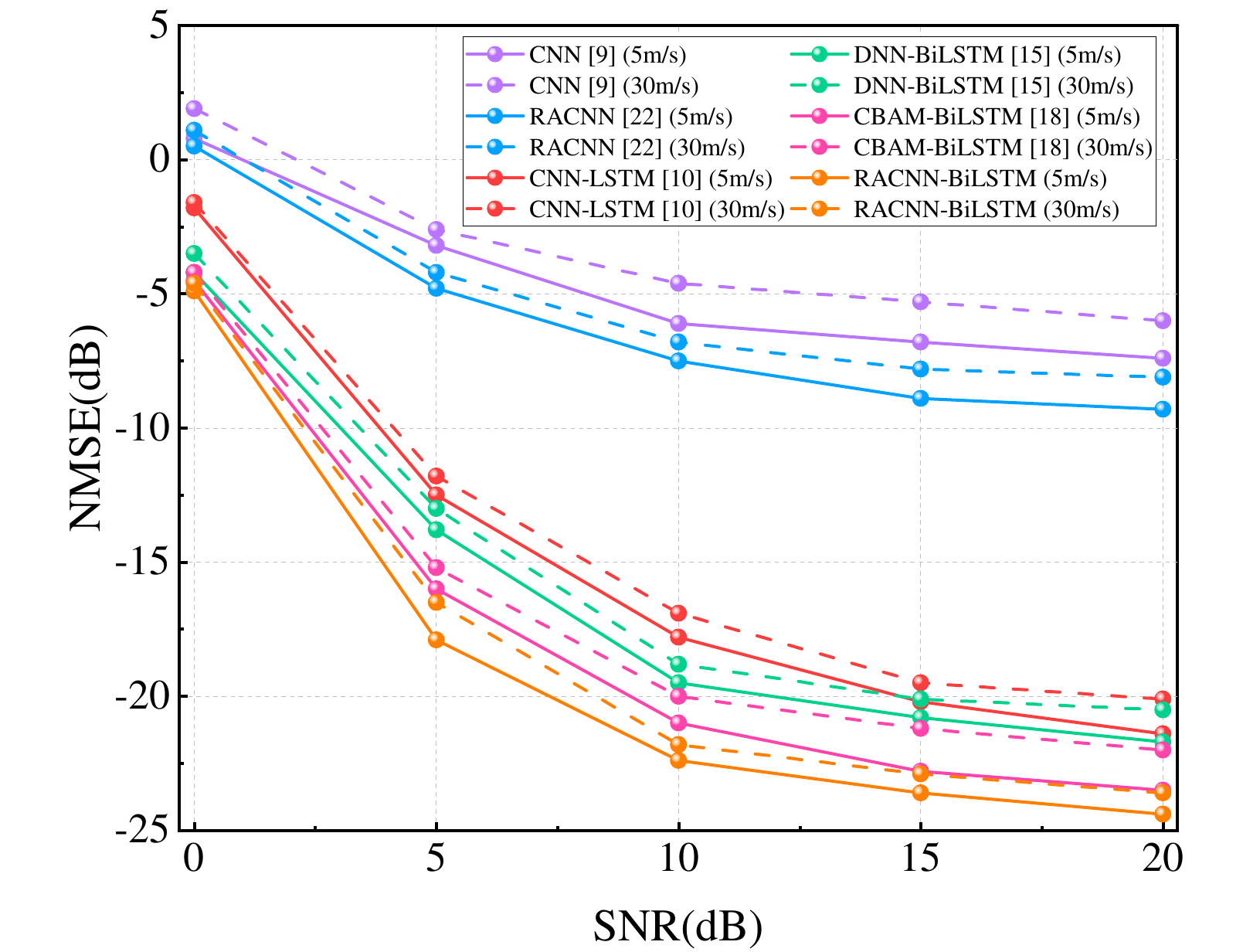}
        \caption{\small User speeds}
        \label{fig:far_speed}
    \end{subfigure}
    \begin{subfigure}{0.32\textwidth}
        \centering
        \includegraphics[width=1\linewidth]{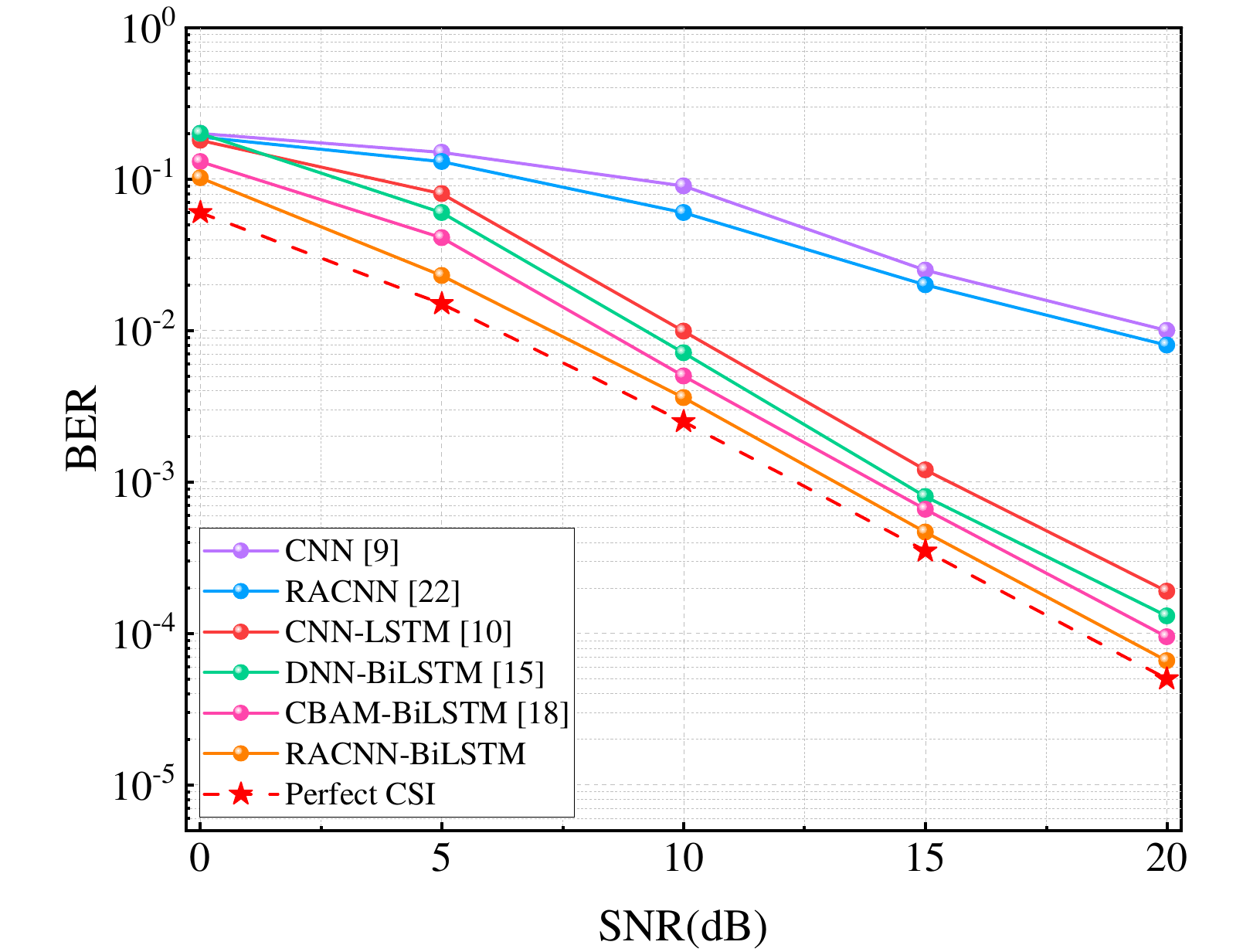}
        \caption{\small BER vs. SNR}
        \label{fig:BER_far}
    \end{subfigure}
    \caption{far-field: Comparison of the performance of different algorithms with respect to the number of transmit antennas, different user speeds ($5\text{m/s}$ and $30\text{m/s}$), and the BER and NMSE performance. }
    \label{fig:(6)}
\end{figure*}

\begin{figure*}[!ht]
    \centering
    \begin{subfigure}{0.32\textwidth}
        \centering
        \includegraphics[width=1\linewidth]{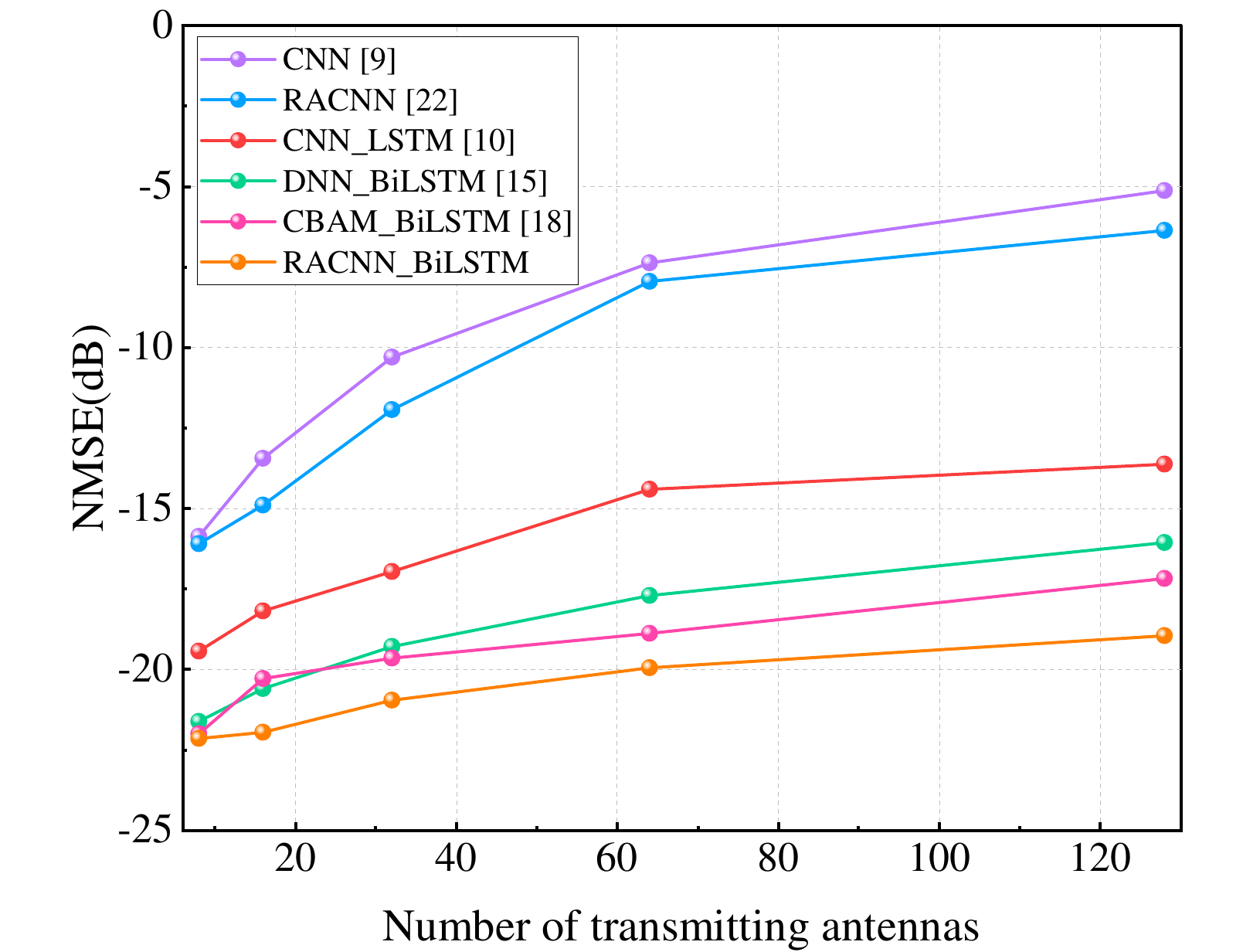}
        \caption{\small Number of transmitting antennas}
        \label{fig:Nt_near}
    \end{subfigure}
    \begin{subfigure}{0.32\textwidth}
        \centering
        \includegraphics[width=1\linewidth]{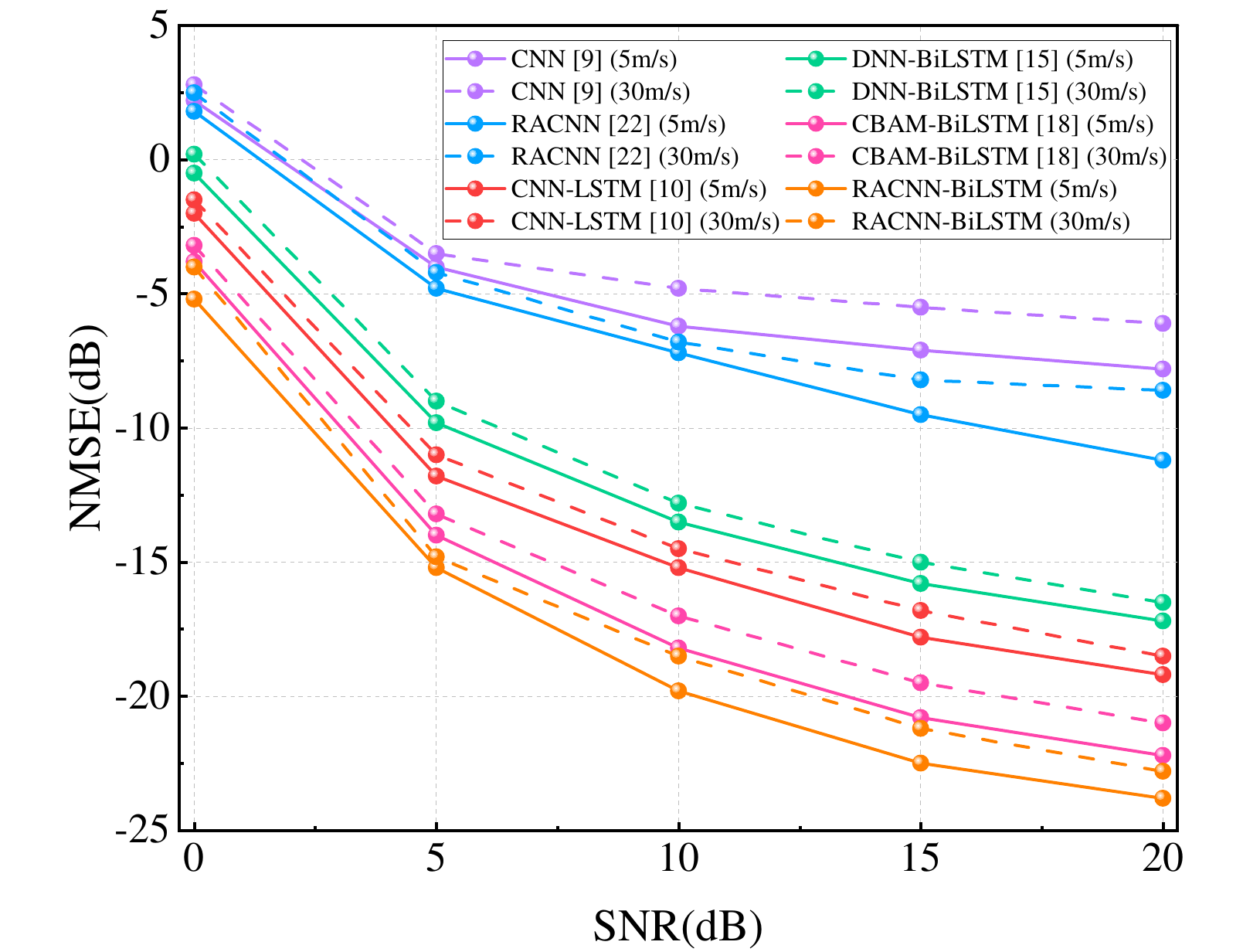}
        \caption{\small User speeds}
        \label{fig:near_speed}
    \end{subfigure}
    \begin{subfigure}{0.32\textwidth}
        \centering
        \includegraphics[width=1\linewidth]{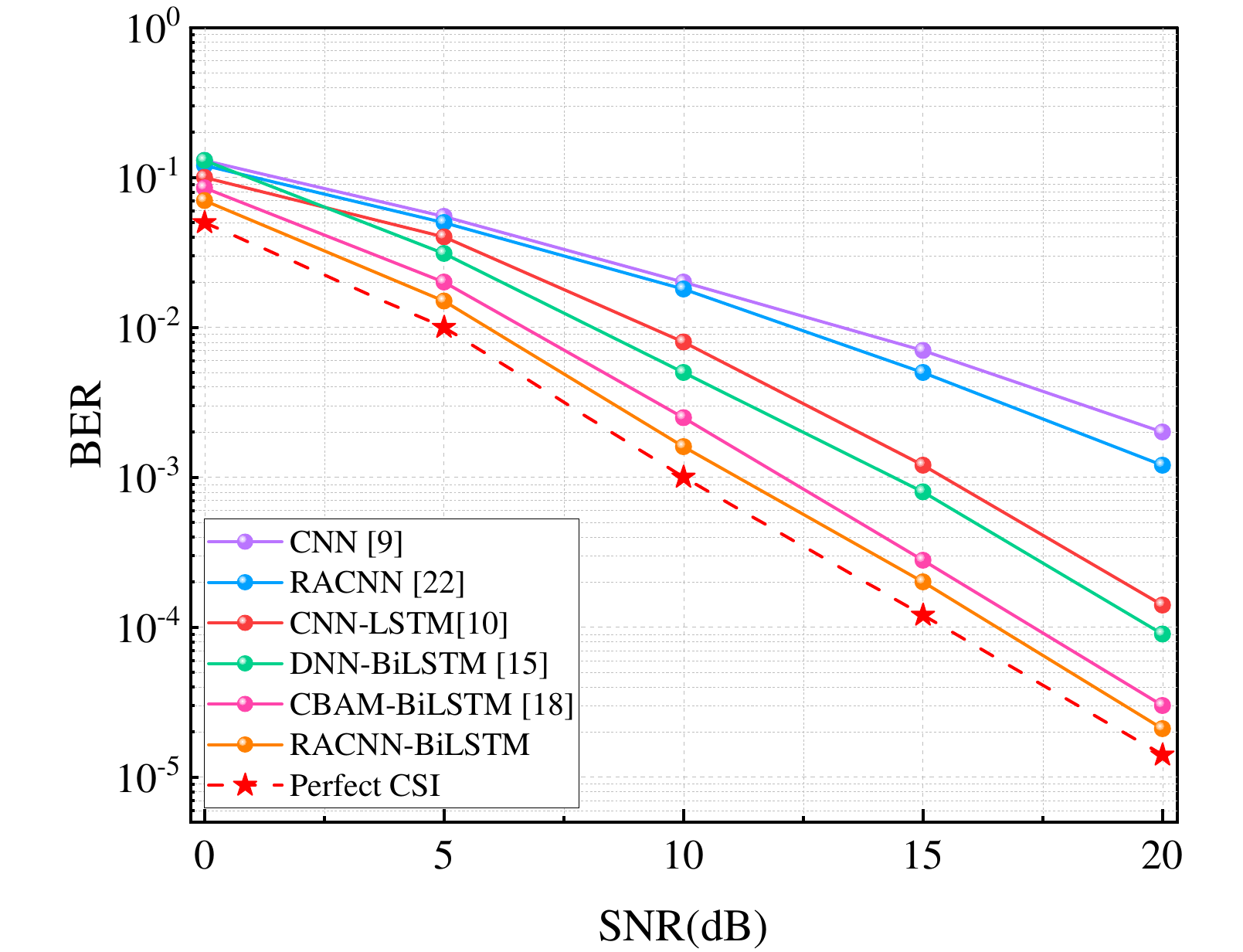}
        \caption{\small BER vs. SNR}
        \label{fig:BER_near}
    \end{subfigure}
     \caption{\small near-field: Comparison of the performance of different algorithms with respect to the number of transmit antennas, different user speeds ($5\text{m/s}$ and $30\text{m/s}$), and the BER and NMSE performance. }
    \label{fig:(7)}
\end{figure*}

In Fig. \ref{fig:Nt_far} and Fig. \ref{fig:Nt_near}, we present the NMSE performance of different algorithms under varying numbers of transmit antennas \{8, 16, 32, 64, 128\} in near-far field scenario, with a fixed SNR of 10 dB. As shown in the figure, the NMSE of channel estimation gradually increases with the number of transmit antennas. This is because increasing the number of antennas requires the model to estimate more channel coefficients, leading to higher errors and increased computational complexity. However, the proposed RACNN-BiLSTM algorithm consistently achieves the lowest NMSE, demonstrating its effectiveness in channel estimation even under higher complexity.

Fig. \ref{fig:far_speed} and Fig. \ref{fig:near_speed} compare the performance of different channel estimation algorithms under near-far field at various UE's moving speeds. As the SNR increases, the NMSE for all algorithms exhibits a monotonic decrease, confirming the robustness of the models under reduced noise conditions. At the same SNR, the curve for $5\text{m/s}$ shows a lower NMSE, indicating that relatively slower user speeds lead to stronger correlation between historical channel states, making it easier to leverage this information to improve estimation accuracy. In contrast, the rapid temporal variation at $30\text{m/s}$ leads to a decrease in the effectiveness of temporal information. Our proposed RACNN-BiLSTM algorithm performs the best, followed closely by CBAM-BiLSTM, highlighting the advantages of the attention mechanism and temporal modeling. CNN-LSTM and DNN-BiLSTM exhibit mid-range performance, while CNN and RACNN, which do not incorporate temporal modeling, perform the worst, underscoring the importance of temporal modeling in dynamic channel estimation.

The bit error rate (BER) performance is an important indicator for evaluating how a channel estimation method impacts the overall system performance. Fig. \ref{fig:BER_far} and Fig. \ref{fig:BER_near} compare the BER performance of different algorithms with user speed of $v_k=\text{10} \text{m}/\text{s}$ in near-far field scenarios. As the SNR increases, the BER of all models gradually decreases, demonstrating the significant impact of signal quality on model performance. Among them, the proposed RACNN-BiLSTM model exhibits the BER performance closest to that of perfect CSI, indicating its superior adaptability and accuracy in complex channel environments. RACNN-BiLSTM achieves an average performance gain of 5.0 dB over CNN and 1.7 dB over CBAM-BiLSTM across the tested SNR range. 

\begin{figure*}[!ht]
    \centering
    \begin{subfigure}{0.32\textwidth}
        \centering
        \includegraphics[width=1\linewidth]{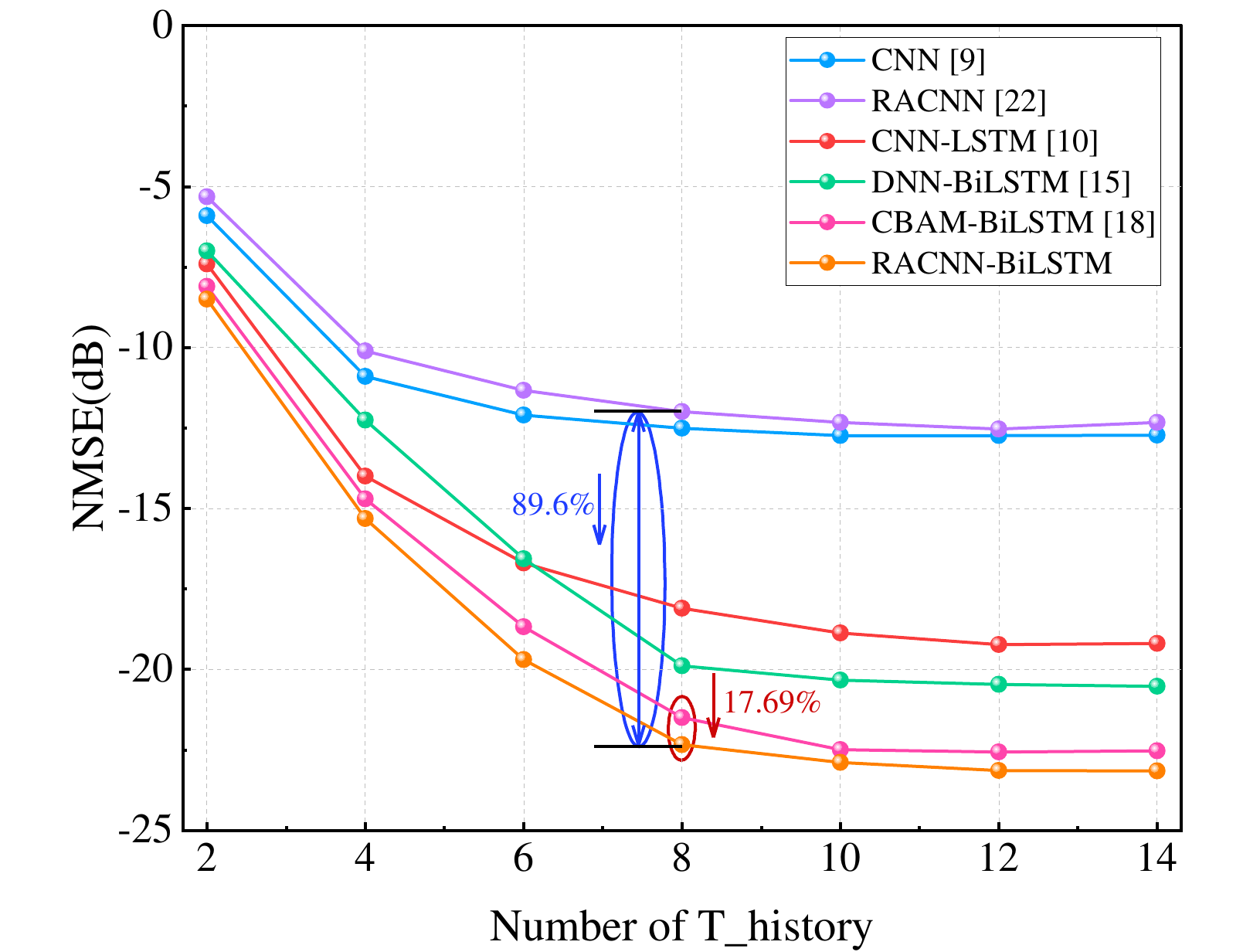}
        \caption{\small T\_history}
        \label{fig:ff_T_hist}
    \end{subfigure}
    \begin{subfigure}{0.32\textwidth}
        \centering
        \includegraphics[width=1\linewidth]{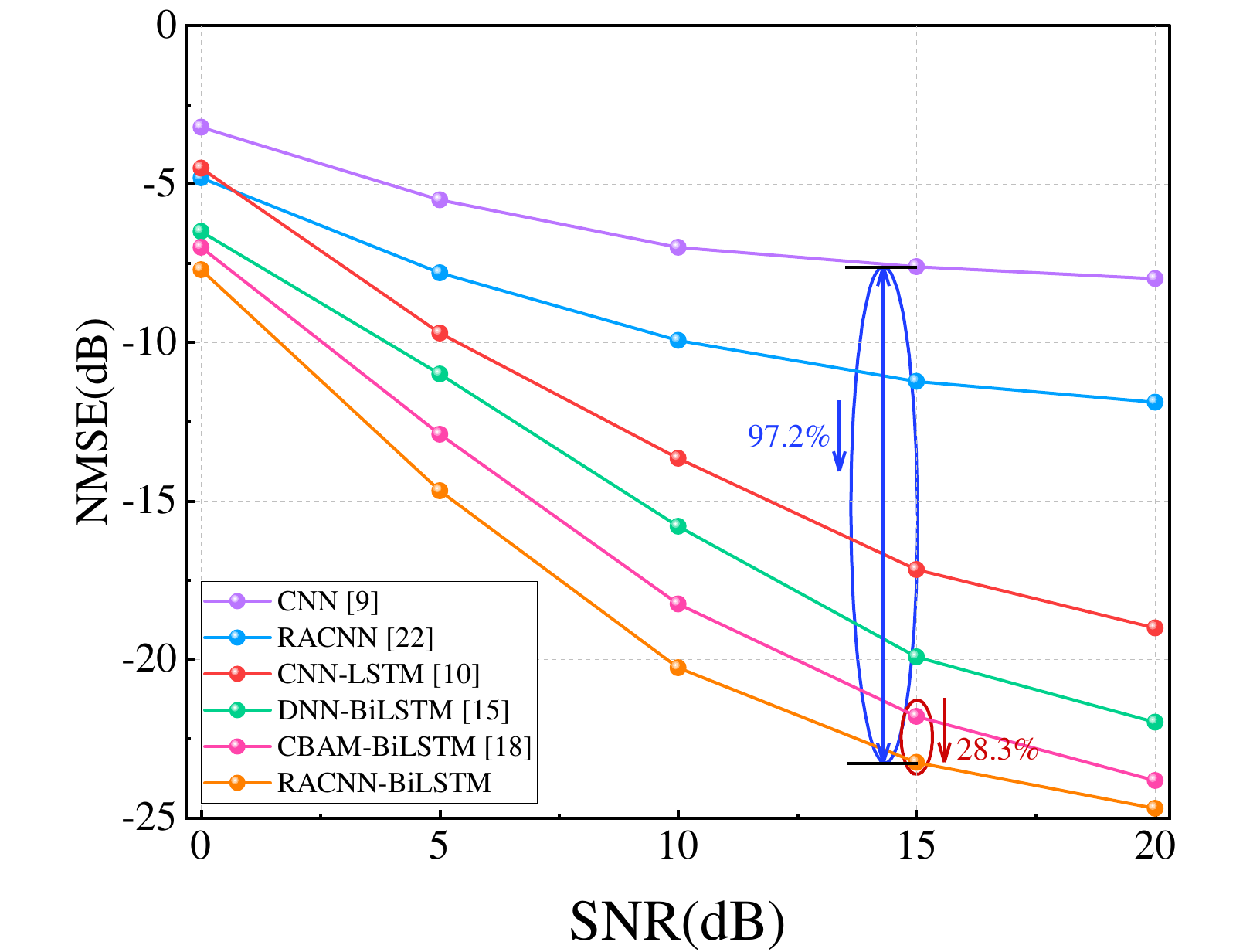}
        \caption{\small $L_f$ = 4}
        \label{fig:ff_Lf=4}
    \end{subfigure}
    \begin{subfigure}{0.32\textwidth}
        \centering
        \includegraphics[width=1\linewidth]{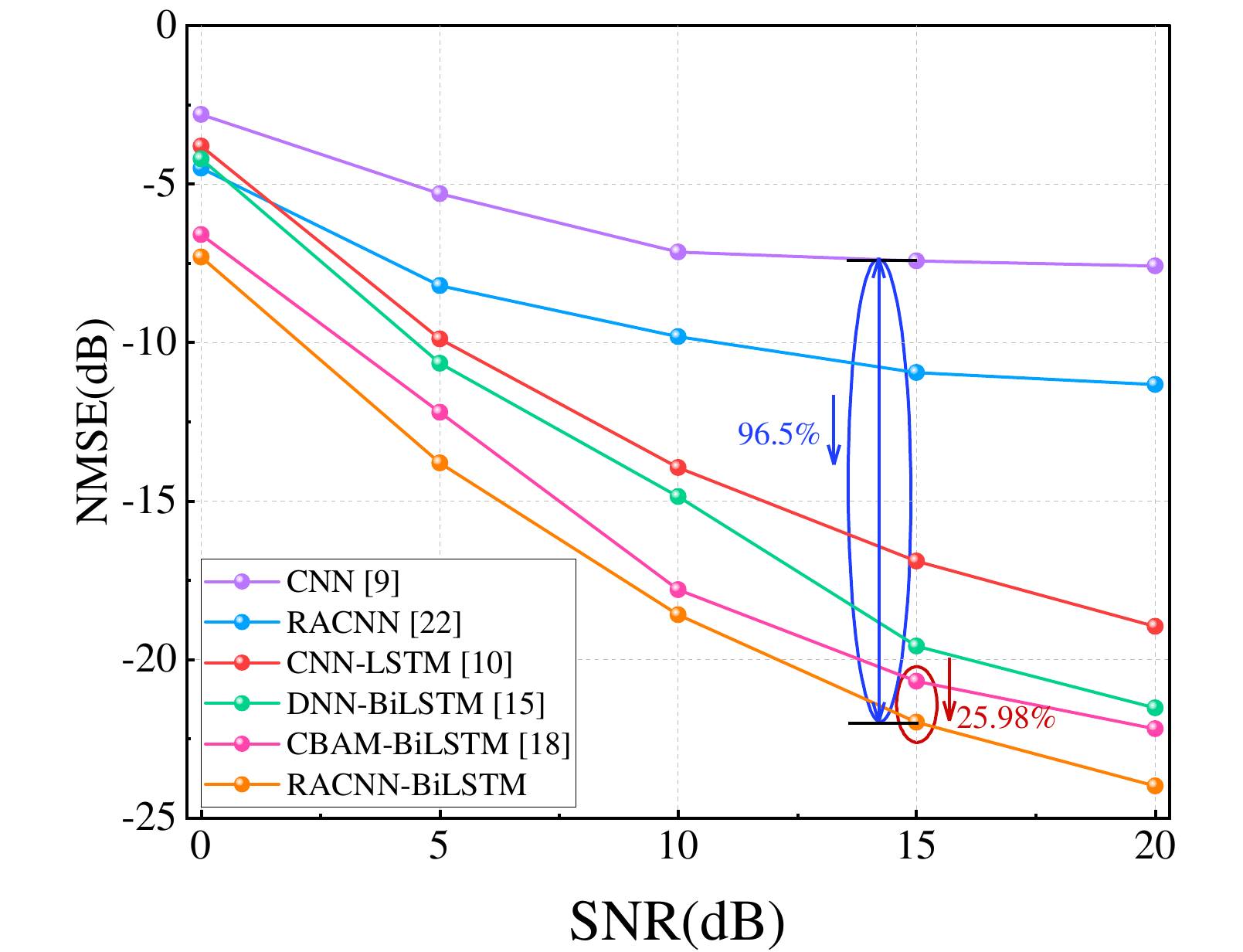}
        \caption{\small $L_f$ = 8}
        \label{fig:ff_Lf=8}
    \end{subfigure}
    \caption{\small far-field: Comparison of performance of different algorithms with respect to the number of T\_history and the number of paths.}
    \label{fig:(8)}
\end{figure*}
\begin{figure*}[!ht]
    \centering
    \begin{subfigure}{0.32\textwidth}
        \centering
        \includegraphics[width=1\linewidth]{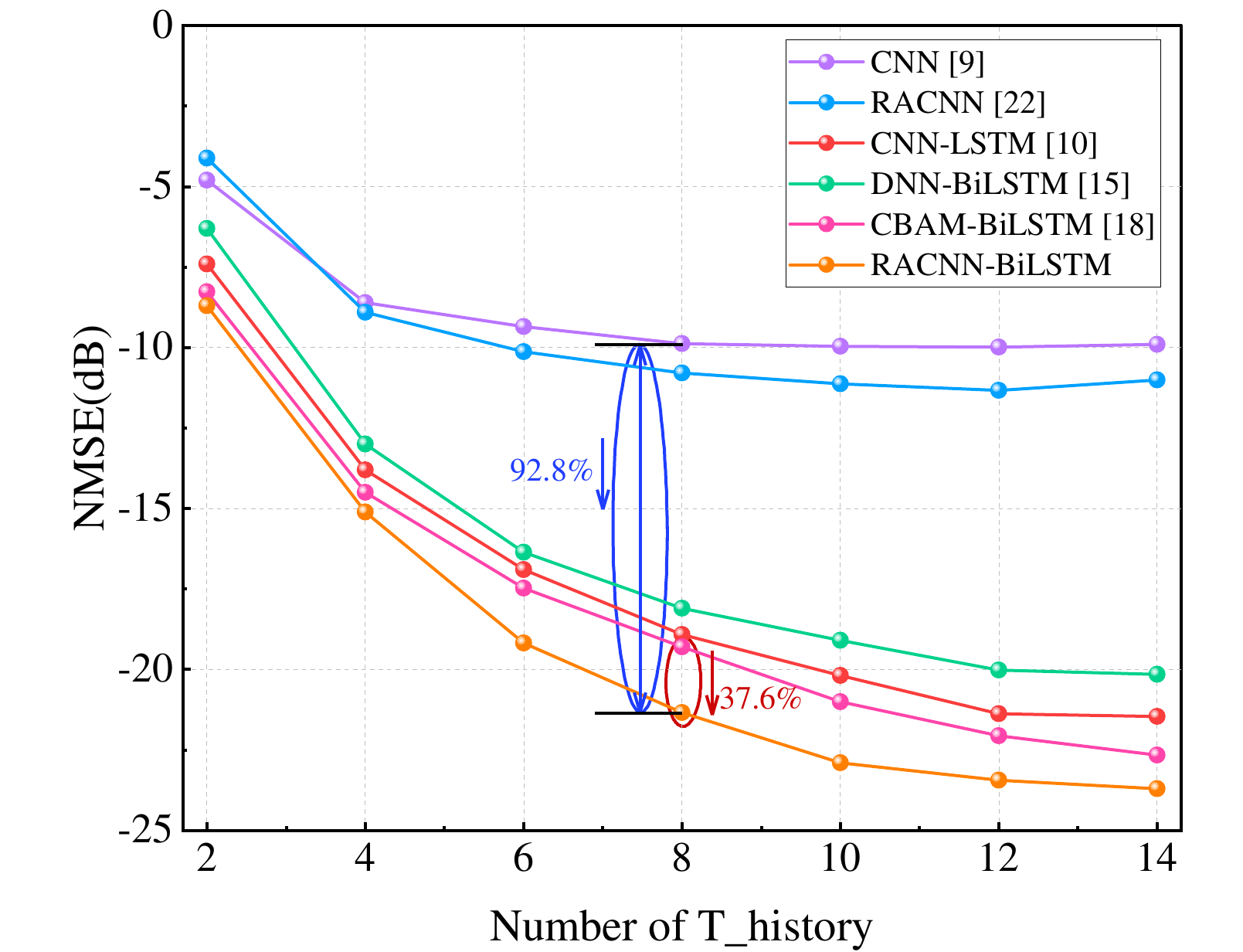}
        \caption{\small T\_history}
        \label{fig:nf_T_hist}
    \end{subfigure}
    \begin{subfigure}{0.32\textwidth}
        \centering
        \includegraphics[width=1\linewidth]{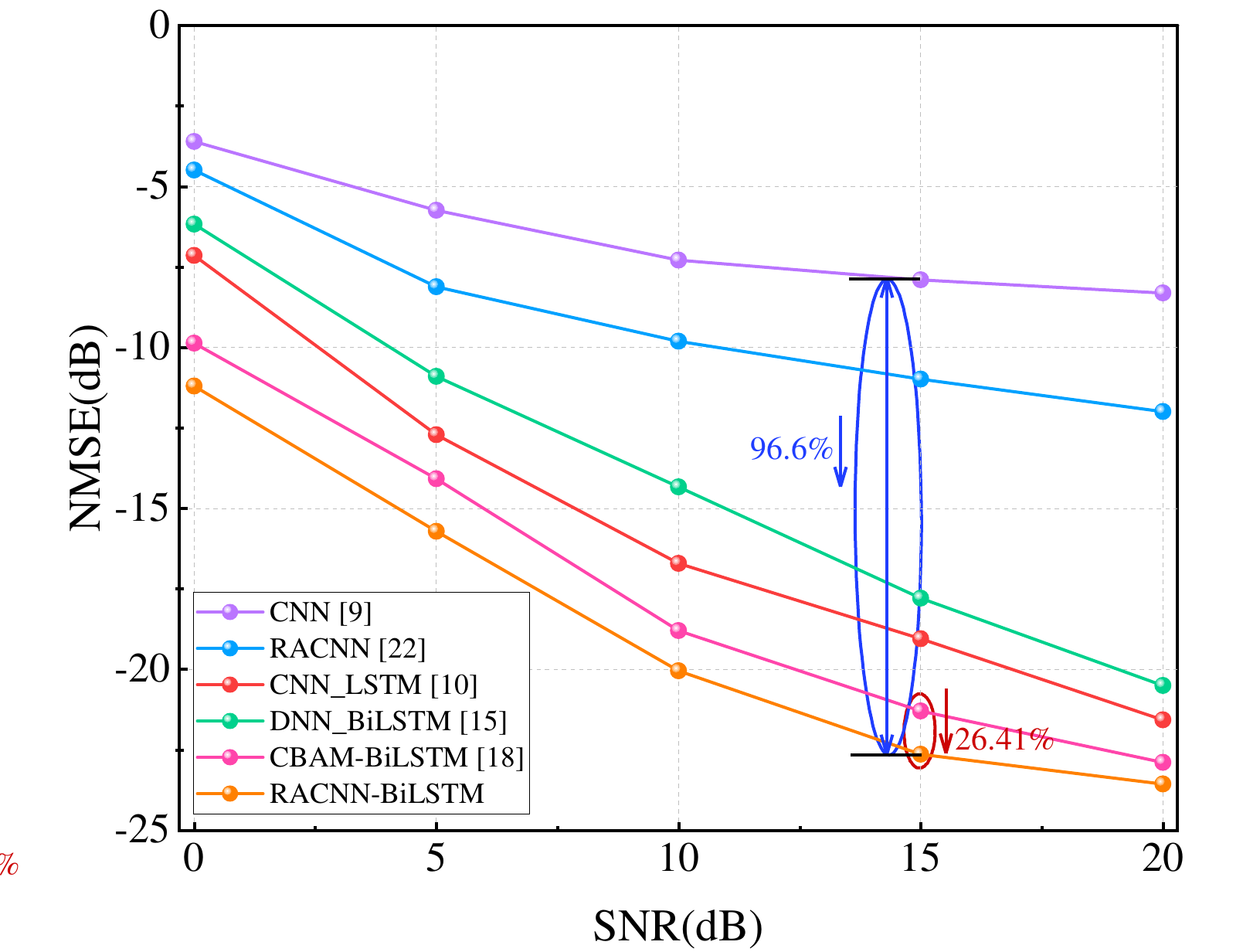}
        \caption{\small $L_n$ = 4}
        \label{fig:nf_Ln=4}
    \end{subfigure}
    \begin{subfigure}{0.32\textwidth}
        \centering
        \includegraphics[width=1\linewidth]{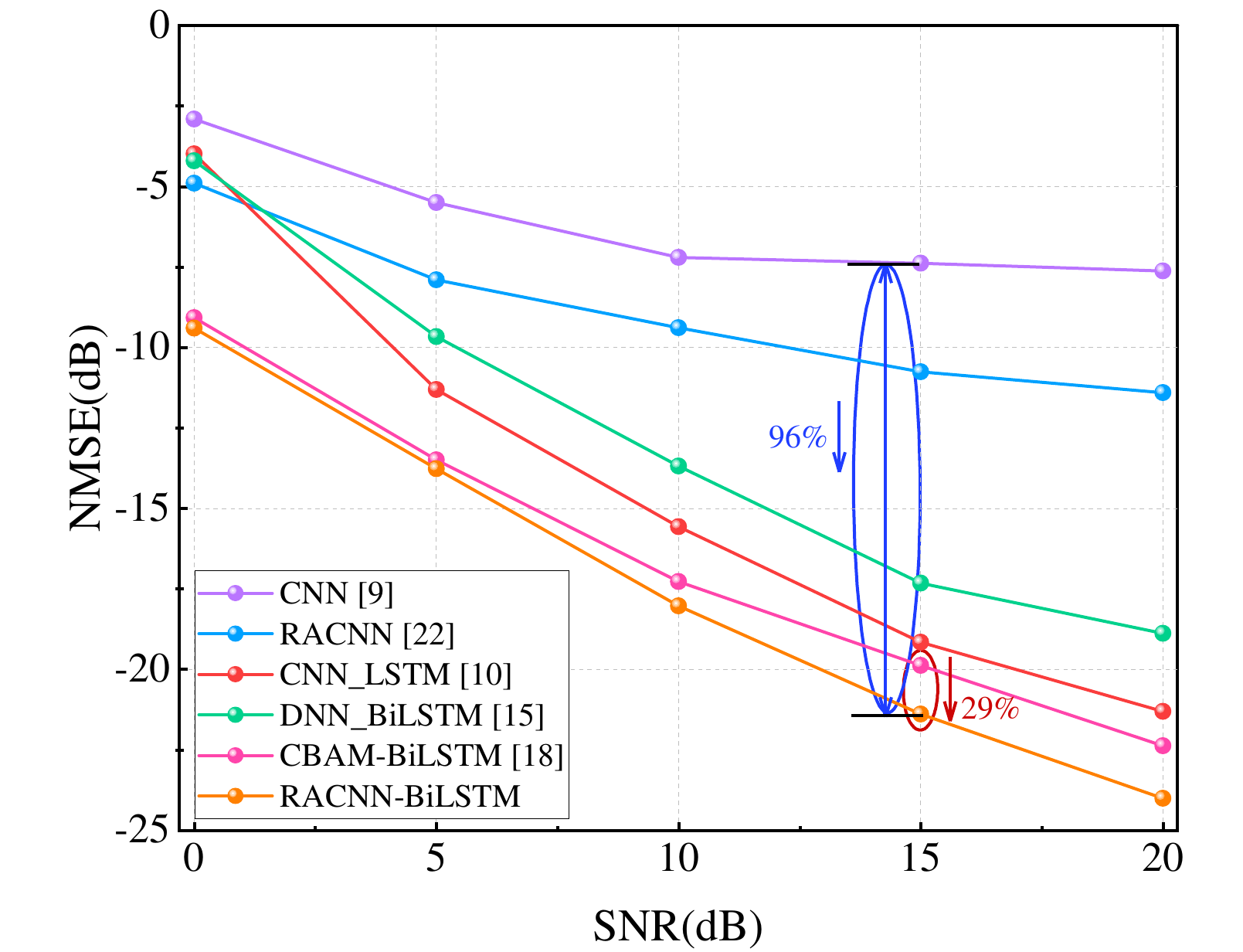}
        \caption{\small $L_n$ = 8}
        \label{fig:nf_Ln=8}
    \end{subfigure}
     \caption{\small near-field: Comparison of performance of different algorithms with respect to the number of T\_history and the number of paths.}
    \label{fig:(9)}
\end{figure*}

The number of historical time slots refers to the quantity of past time slots utilized by the model in channel estimation, which determines the maximum temporal dependency range that the model can capture. Fig. \ref{fig:ff_T_hist} and Fig. \ref{fig:nf_T_hist} illustrate the performance comparison of different algorithms for channel estimation with varying historical time slot lengths, when the SNR is set to 10 dB and the user speed is \text{10}\,\text{m/s}. When the number of historical time slots increases from 2 to 6, the performance of all models improves significantly. This indicates that as the number of historical time slots increases, the models can capture more prior information about the channel state, allowing them to better identify temporal variations in the channel and improve the accuracy of channel estimation. In the range of 6 to 10 historical time slots, the rate of NMSE decrease slows down significantly, and performance improvement tends to plateau. This is because, in a dynamic environment, as time progresses, channel variations become more pronounced, and information from earlier historical time slots may no longer effectively reflect the current channel state. When the number of historical time slots exceeds 10, CNN, RACNN, and CNN-LSTM exhibit slight performance degradation, indicating that an excessive number of historical time slots may cause the model to lag in response and introduce noise, making it unable to adapt to rapid channel changes, while also lacking more effective temporal modeling capabilities. Our proposed RACNN-BiLSTM method continues to show lower NMSE as the number of historical time slots increases. 
When the number of historical time slots is 8, the proposed method achieves an average NMSE reduction of at least 91\% and 27.6\% compared to CNN and CBAM-BiLSTM, respectively, in both near-field and far-field scenarios. These results suggest that in dynamic channel estimation, appropriately utilizing the length of historical information strikes an optimal balance between information gain and computational complexity. However, when the number of historical time slots becomes excessive, the model may face redundancy and overfitting issues, leading to a decline in performance.
   
To investigate the impact of multipath richness on channel estimation performance, we conducted comparative experiments in two scenarios with different path numbers ($L_f=4$ and $L_f=8$). As illustrated in Fig. \ref{fig:ff_Lf=4}, \ref{fig:ff_Lf=8}, \ref{fig:nf_Ln=4}, \ref{fig:nf_Ln=8}, the NMSE for $L_f=4$ is on average 1.2dB lower than for $L_f=8$ across the whole SNR range, indicating that when the number of paths is higher, the channel environment becomes more complex, potentially leading to higher estimation errors due to path interference and signal distortion, resulting in a higher NMSE. Among all schemes, RACNN-BiLSTM and CBAM-BiLSTM yield the lowest errors, confirming the benefit of combining attention mechanisms with BiLSTM. Moreover, the consistent superiority of RACNN-BiLSTM demonstrates that its multi-head self-attention, by modeling long-range dependencies in parallel subspaces, better captures the latent time-frequency structures of dense multipath channels than the sequential channel-spatial attention used in CBAM, leading to higher estimation accuracy and stronger robustness.

\section{Conclusion}\label{sec:conclusion}
In this paper, we propose a DL channel estimation method aimed at addressing the dynamic channel estimation problem in UAV-assisted communication systems. This method integrates CNNs for spatial feature extraction, BiLSTM for temporal modeling, and introduces a residual self-attention mechanism to enhance the representation of key channel features. Furthermore, integrating the location information of both communication parties into the network facilitates the model's ability to effectively differentiate between the characteristics of near-field and far-field environments, thereby enhancing the accuracy of channel estimation. Extensive simulations demonstrate that the proposed RACNN-BiLSTM model exhibits superior estimation accuracy under both near-field and far-field conditions, benefiting from the advantages of location information and the attention mechanism. Compared to other schemes, the location-based channel estimation method improves performance by at least 36.3\% and 24.2\% in far-field and near-field scenarios, respectively, enabling better adaptation to dynamic environmental changes and enhancing channel estimation accuracy.

\bibliographystyle{IEEEtran}
\bibliography{ref}
\end{document}